\pdfoutput=1
\documentclass{jfm}

\usepackage{subfigure}

\usepackage{graphicx}
\usepackage[framed,numbered,autolinebreaks,useliterate]{mcode}
\usepackage{natbib}
\usepackage{amsmath}
\usepackage{bm}
\usepackage{float} 
\usepackage{graphicx}
\usepackage{rotating}
\usepackage{latexsym}
\usepackage{tikz}
\usepackage{pgfplots}
\usepackage{algorithm}
\usepackage{algorithmic}
\pgfplotsset{
    ylabel style={rotate=-90},
    tick label style={font=\small},
    label style={font=\small},
    legend style={font=\footnotesize},
    legend style={
        at={(1.03,0.5)},
        anchor=west,
        legend columns=1,
        cells={anchor=west},
        font=\footnotesize,
        rounded corners=1pt,
    },
}

\ifCUPmtlplainloaded \else
  \checkfont{eurm10}
  \iffontfound
    \IfFileExists{upmath.sty}
      {\typeout{^^JFound AMS Euler Roman fonts on the system,
                   using the 'upmath' package.^^J}%
       \usepackage{upmath}}
      {\typeout{^^JFound AMS Euler Roman fonts on the system, but you
                   dont seem to have the}%
       \typeout{'upmath' package installed. JFM.cls can take advantage
                 of these fonts,^^Jif you use 'upmath' package.^^J}%
      }
  \else
  \fi
\fi

\ifCUPmtlplainloaded \else
  \checkfont{msam10}
  \iffontfound
    \IfFileExists{amssymb.sty}
      {\typeout{^^JFound AMS Symbol fonts on the system, using the
                'amssymb' package.^^J}%
       \usepackage{amssymb}%
         \let\leq=\leqslant
       \let\ge=\geqslant  \let\geq=\geqslant
      }{}
  \fi
\fi

\ifCUPmtlplainloaded \else
  \IfFileExists{amsbsy.sty}
    {\typeout{^^JFound the 'amsbsy' package on the system, using it.^^J}%
     \usepackage{amsbsy}}
    {\providecommand\boldsymbol[1]{\mbox{\boldmath $##1$}}}
\fi

\newsavebox{\astrutbox}
\sbox{\astrutbox}{\rule[-5pt]{0pt}{20pt}}

\title[A New Approach to MOR of the Navier-Stokes]{A Novel Model Order Reduction Approach for Navier-Stokes Equations at High Reynolds Number}

\author[Maciej J. Balajewicz and Earl H. Dowell]%
{M\ls A\ls C\ls I\ls E\ls J\ns J.\ns B\ls A\ls L\ls A\ls J\ls E\ls W\ls I\ls C\ls Z$^1$%
  \thanks{Email address for correspondence: maciej.balajewicz@stanford.edu},\ns
E\ls A\ls R\ls L\ns H.\ns D\ls O\ls W\ls E\ls L\ls L$^2$\break
\and B\ls E\ls R\ls N\ls D\ns R.\ns N\ls O\ls A\ls C\ls K$^3$}

\affiliation{$^1$Department of Aeronautics and Astronautics, Stanford University, Stanford, CA 94305, USA\\
$^2$Department of Mechanical Engineering and Materials Science, Duke University,
Durham, NC 27708, USA\\
$^3$Institut PPRIME, CNRS -- Universit\'e de Poitiers -- ENSMA, UPR 3346,
           D\'epartment Fluides, Thermique, Combustion,
           CEAT, 43 rue de l'A\'erodrome, F-86036 POITIERS Cedex, France}

\pubyear{2010}
\volume{650}
\pagerange{119--126}
\begin{document}

\maketitle

\begin{abstract}
A new approach to model order reduction of the Navier-Stokes equations at high Reynolds number is proposed. Unlike traditional approaches, this method does not rely on empirical turbulence modeling or modification of the Navier-Stokes equations. It provides spatial basis functions different from the usual proper orthogonal decomposition basis function in that, in addition to optimally representing the training data set, the new basis functions also provide stable and accurate reduced-order models. The proposed approach is illustrated with two test cases: two-dimensional flow inside a square lid-driven cavity and a two-dimensional mixing layer.
\end{abstract}


\section{Introduction}
\indent Numerical simulation of fluid flows is usually a very computationally intensive endeavor. Even when adequate computational resources are available, numerical simulations often provide too little understanding of the solutions they produce. There are significant scientific and engineering benefits in developing and studying Model Order Reduction (MOR) techniques capable of producing Reduced Order Models (ROMs) of complex fluid flows that retain physical fidelity while substantially reducing the size and cost of the computational model. \\
\indent In fluid flow applications, Proper Orthogonal Decomposition (POD) and Galerkin projection form a popular MOR strategy~\citep{Noack:2011,Holmes:2012}. Despite its popularity, deriving robust and accurate ROMs using the POD-Galerkin approach remains challenging and many important issues remain unresolved. In particular, application of the POD-Galerkin strategy to turbulent fluid flows remains an active area of research. Turbulence is a flow regime characterized by chaotic, multi-scale dynamics in both space and time. In the limit of high Re numbers, the dynamics of turbulence are qualitatively universal: large scale flow features are broken down into smaller and smaller scales until the scales are fine enough that viscous forces can dissipate their energy~\citep{Pope:2000,Moin:1998,Tennekes:1972}. Application of the standard POD-Galerkin strategy to a turbulent flow is problematic because POD, by construction, is biased toward the large, energy containing scales of the turbulent flow. ROMs generated using only the first most energetic POD basis functions are, therefore, not endowed with the natural energy dissipation of the smaller, lower energy turbulent scales. Currently, two broad strategies for adapting the POD-Galerkin approach to high Re number flows are available: (1) direct modeling of small scale dynamics by inclusion of a sufficiently large number of POD basis functions and (2) approximate modeling of small scale dynamics using empirical turbulence models. Unfortunately, neither strategies is ideal. Inclusion of a large number of POD basis functions creates very large POD-Galerkin ROMs that are still very computationally expensive to solve. Addition of turbulence models is equally undesirable because the empirical terms modify the dynamics of the Navier-Stokes equations~\citep{Rempfer:1994,Aubry:1988,Ukeiley:2001,Sirisup:2004,Iliescu:2012,Bailon:2012}. \\
\indent In this paper, a new approach to MOR of turbulent Navier-Stokes fluid flows is proposed. Instead of modeling the small, energy-dissipative scales using empirical eddy viscosity models, the small scales are modeled directly using spatial basis functions that are different from the standard POD basis functions. In addition to optimally representing the solution, these new basis functions also resolve a greater proportion of the small, energy-dissipative scales of the turbulent flow. The proposed methodology is formulated as a small-scale constrained minimization problem that can be solved numerically using standard, off-the-shelf MATLAB algorithms. \\
\indent This paper is organized as follows. In \S 2 the POD and Galerkin methods are summarized and the approach is applied to the incompressible Navier-Stokes equations. In \S 3 the proposed new approach is introduced and the algorithm and its numerical implementation are outlined. In \S 4 two classical benchmarks, the lid-driven cavity and a mixing layer are modeled using the standard POD-Galerkin approach and the new approach. Finally, in \S 5, the main results are summarized and future prospects laid out.
\section{Methodology} \label{sec:Methodology}
\subsection{Spectral methods}
\label{sec:Spectral methods}
Consider a dynamical system that evolves in a Hilbert space $H$, $\bm{u}(t) \in H$ governed by
\begin{equation}
    \bm{\dot u} = X\left( \bm{u} \right)
\end{equation}
where $X$ is a vector field on $H$. Often, the governing variable $\bm{u}$ is spatio-temporal, $\bm{u}=\bm{u}(\bm{x},t)$, $\bm{x} \in \Omega$, $t \in \left[ {0,T} \right]$ and thus, $X$ is a spatial differential operator. In the spectral approach~\citep{Boyd:2001,Canuto:1991,Canuto:2006}, the governing variable, $\bm{u}(\bm{x},t)$ is discredited using separable basis functions, $a_i(t)$ and $\bm{u}_i (\bm{x})$
\begin{equation}
    \bm{u}(\bm{x},t) \approx \bm{u}^{[1, \ldots ,n]}  \equiv \sum\limits_{i}^n {a_i (t) \bm{u}_i (\bm{x})}
\end{equation}
where $\left\{ {\left. {\bm{u}_i  \in H} \right|i = 1, \ldots ,n} \right\}$ is a basis for the subspace of $H$. In the method of lines, the spatial basis functions, $\bm{u}_i$ are known a priori and the goal, therefore, is to find temporal coefficients, $a_i$ that satisfy the differential equation. In general, the origin and/or form of the spatial basis functions $\bm{u}_i$ is quite arbitrary. In the context of spectral methods in Computational Fluid Dynamics (CFD) the spatial basis functions, $\bm{u}_i$ are usually analytical functions such as trigonometric functions and power functions, i.e. Chebyshev polynomials. The benefit of using these functions is that their spatial derivatives are known and numerically efficient algorithms such as the Fast Fourier Transform (FFT) can be utilized. In the context of MOR, the spatial basis functions can be derived a priori using completeness conditions~\citep{Ladyzhenskaia1:1969,Noack:1994}, from Navier-Stokes eigenfunctions~\citep{Joseph:1976}, or a posteriori from a snapshot of a solution data set, like the Proper Orthogonal Decomposition (POD)~\citep{Holmes:2012} or Dynamic Mode Decomposition (DMD)~\citep{Rowley:2009,Schmid:2010}. \\
\indent The temporal coefficients, $a_i$ of the spectral discretized dynamical system are chosen such that the error is orthogonal to a subspace of $H$. Let $\left\{ {\left. {\bm{v}_i  \in H} \right|i = 1, \ldots ,n} \right\}$ be a basis for a subspace of $H$. We seek $a_i$ such that
\begin{equation}
    \left\langle {\bm{v}_i ,\bm{\dot u}^{[1, \ldots ,n]} } \right\rangle _\Omega   - \left\langle {\bm{v}_i ,X\left( {\bm{u}^{[1, \ldots ,n]} } \right)} \right\rangle _\Omega   = 0
\end{equation}
where $\left\langle { \cdot , \cdot } \right\rangle_\Omega$ is an appropriately defined spatial inner product. In the Galerkin projection approach, the trail basis are equal to the test basis; $\bm{v}_i  \equiv  \bm{u}_i$. The projection yields a set of evolution equations for the temporal coefficients $a_i$
\begin{equation}
    \dot a_i  = f_i (a)
\end{equation}
Given a set of initial conditions the evolution equation can be integrated using standard numerical integration techniques.
\subsection{Proper orthogonal decomposition (POD)} \label{sec:Proper orthogonal decomposition (POD)}
The Proper Orthogonal Decomposition (POD) is an information compression technique that is applicable to a wide variety of problems including digital image compression~\citep{Richards:2006}, signal processing~\citep{Caccaro:1991} and bioinformatics~\citep{Wall:2003}. We express the POD optimality condition in terms of the turbulent kinetic energy, $e(t)$ of the flow defined as
\begin{equation}
    \label{eqn:tke_full}
    e(t) \equiv \frac{1}{2}\int\limits_\Omega  {\left| {\bm{u}(\bm{x},t)} \right|^2 dx}
\end{equation}
POD provides spatial and temporal basis functions whose turbulent kinetic energy
\begin{equation}
    e^{[1, \ldots ,n]} (t) \equiv \frac{1}{2}\int\limits_\Omega  {\left| {\sum\limits_{i = 1}^n {a (t) \bm{u} (\bm{x})} } \right|^2 dx}
\end{equation}
is as close as possible, on average, to the turbulent kinetic energy, $e(t)$ of a solution snapshot of the Navier-Stokes, $\bm{u}_s (\bm{x},t)$. In other words, the POD optimality condition can be formulated as follows
\begin{equation}
\label{Eqn:POD optimization problem}
    \begin{aligned}
    & \underset{a,\bm{u}} {\operatorname{arg\,min}}
    & & \displaystyle \left\langle {\int\limits_\Omega  {\left| {\bm{u}_s (\bm{x},t) - \sum\limits_{i = 1}^n {a_i(t) \bm{u}_i(\bm{x})} } \right|^2 dx} } \right\rangle _T  \\
    & \text{s.t.}
    & & (1) \hspace{0.1mm} \left\langle {\bm{u}_i ,\bm{u}_j } \right\rangle _\Omega   = \int_\Omega  {\bm{u}_i(\bm{x})  \cdot \bm{u}_j(\bm{x}) } d\bm{x} = \delta _{ij} \\
    &&& (2) \hspace{0.1mm} \left\langle {a_i ,a_j } \right\rangle _T  = \frac{1}{T}\int_T {a_i (t)a_j (t)dt}  = \lambda _{ij}
    \end{aligned}
\end{equation}
where $\delta_{ij}$ is the Kronecker delta and the diagonal values of the correlation matrix, $\lambda$ are the POD eigenvalues. The discretized equivalent of Eq.~\eqref{Eqn:POD optimization problem} is solved using the well known Eckart-Young theorem~\citep{Demmel:1997} of Singular Value Decomposition (SVD) and the method of snapshots~\citep{Sirovich:1987}.
\subsection{Galerkin MOR of the Navier Stokes equation} \label{sec:Projection-based ROMs of the Navier Stokes equation}
The evolution of the velocity field, $\bm{u}(\bm{x},t)$ of an incompressible, isentropic, Newtonian fluid is governed by the following Navier-Stokes equations
\begin{subequations}
    \begin{align}
            \nabla  \cdot {\bm{u}}  &= 0 \\
            \displaystyle \frac{{\partial {\bm{u}}}}{{\partial t}} &= \nu \Delta {\bm{u}} - \nabla  \cdot ( {\bm{u}} {\bm{u}}) - \nabla p
    \end{align}
\end{subequations}
where $\nu$ is the viscosity and $p$ is the pressure. In the standard Galerkin MOR approach, the fluid velocity vector, $\bm{u}(\bm{x},t)$ is discretized spectrally using separable spatial, $\bm{u}_i(\bm{x})$ and temporal, $a_i(t)$ basis functions
\begin{equation}
    \label{Eqn:spectral approximation in velocity}
    \bm{u}(\bm{x},t) \approx \bm{u}^{[0,\cdots,n]}   = \bm{u}_0 + \sum\limits_{i=1}^n a_i (t) {\bm{u}_i (\bm{x})}
\end{equation}
where ${\bm{u}_0}$ is the temporal mean. Galerkin projection yields a set of $n$ coupled, quadratic Ordinary Differential Equations (ODEs) in the following canonical form
\begin{equation}
    \label{Eqn:GS_VTE}
    \dot a_i  = \sum\limits_{j,k = 1}^n {Q_{ijk} a_j a_k }  + \sum\limits_{j = 1}^n {L_{ij} a_j  + b_i } {\rm{ }}
\end{equation}
For divergence-free spatial basis functions and steady dirichlet boundary conditions, the Galerkin matrices are of the form
\begin{subequations}
    \begin{align}
        \displaystyle Q_{ijk} &= \left\langle {\bm{u}_i ,\nabla \cdot(\bm{u}_j \bm{u}_k )} \right\rangle _\Omega \\
        \displaystyle L_{ij}  &= \left\langle {\bm{u}_i , \nu \Delta \bm{u}_j } \right\rangle _\Omega
            + \left\langle {\bm{u}_i , - \nabla  \cdot (\bm{u}_0 \bm{u}_j )} \right\rangle _\Omega
            + \left\langle {\bm{u}_i , - \nabla  \cdot (\bm{u}_j \bm{u}_0 )} \right\rangle _\Omega\\
        \displaystyle b_{i} &= \left\langle {\bm{u}_i ,\nu \Delta \bm{u}_0  - \nabla \cdot(\bm{u}_0 \bm{u}_0 )} \right\rangle _\Omega
    \end{align}
\end{subequations}
For open flows the pressure term cannot be ignored. In practice however, the pressure term is rarely modeled in full. For example, the contribution from the pressure term can usually be modeled well using a linear fit of the linear Galerkin terms, $L_{ij}$~\citep{Galletti:2004,Noack:2005}.
\section{A Novel, Galerkin-based Model Order Reduction approach for the Navier-Stokes equations at high Reynolds number} \label{sec:Reduced Order Model (ROM) stability and accuracy}
Turbulence is a flow regime characterized by chaotic, multi-scale dynamics in both space and time. POD basis functions are by construction, biased towards the large, energy containing scales of the turbulent flow. Galerkin ROMs generated using only the first most energetic POD basis functions are therefore not endowed with the natural energy dissipation of the smaller, lower energy turbulent scales. Since accounting for all the scales of the turbulent flow directly by including a large enough number of POD basis function is computationally prohibitive, the traditional approach as been to model the contribution of the small scales empirically. In this section, a new, non-empirical MOR approach for turbulent Navier-Stokes fluid flows is proposed. Instead of modeling the small, energy-dissipative scales using empirical eddy viscosity models, the small scales are modeled directly using spatial basis functions that are different from the standard POD basis functions. \\
\indent The productive/dissipative characteristics of a set of basis functions can be quantified using the time-averaged rate of turbulent kinetic production. The turbulent kinetic energy of a spectrally discretized fluid flow is defined as
\begin{equation}
    \label{Eqn:e}
    e^{[1, \ldots ,n]} (t) \equiv \frac{1}{2}\int\limits_\Omega  {\left| {\sum\limits_{i = 1}^n {a_i (t) \bm{u}_i (\bm{x})} } \right|^2 dx}
\end{equation}
Equation~\eqref{Eqn:e} can be expressed as a function of the temporal coefficients, $a_i(t)$ alone by exploiting the orthonormality of the spatial basis functions
\begin{equation}
    e^{[1, \ldots ,n]} (t) = \frac{1}{2}\sum\limits_{i = 1}^n {a_i(t)^2 }
\end{equation}
A dynamical equation for the turbulent kinetic energy is derived as follows
\begin{equation}
    \label{Eqn:e2}
    \frac{d}{{dt}}e^{[1, \ldots ,n]} (t) = \frac{d}{{dt}}\frac{1}{2}\sum\limits_{i = 1}^n {a_i (t)^2 }  = \sum\limits_{i = 1}^n {a_i (t)\frac{d}{{dt}}a_i (t)}
\end{equation}
Using Eq.~\eqref{Eqn:GS_VTE} together with Eq.~\eqref{Eqn:e2}, we arrive at
\begin{equation}
    \label{Eqn:e3}
   \frac{d}{{dt}}e^{[1, \ldots ,n]} (t)  = \sum\limits_{i,j,k = 1}^n {Q_{ijk} a_i a_j a_k }  + \sum\limits_{i,j = 1}^n {L_{ij} a_i a_j }  + \sum\limits_{i = 1}^n {b_i a_i }
\end{equation}
Finally, taking the time-average of Eq.~\eqref{Eqn:e3}, results in
\begin{equation}
    \label{Eqn:rate_of_e}
    \left\langle {\frac{d}{{dt}}e^{[1, \ldots ,n]} (t) } \right\rangle _T  = \sum\limits_{i,j,k = 1}^n {Q_{ijk} \Upsilon_{ijk} }  + \sum\limits_{i,j = 1}^n {L_{ij} \lambda_{ij}  + } \sum\limits_{i = 1}^n {b_i \mu_i }
\end{equation}
where $\Upsilon \in \mathbb{R}^{n \times n \times n}$, $\lambda \in \mathbb{R}^{n \times n}$ and $\mu \in \mathbb{R}^{n}$ are matrices containing the temporal correlations
\begin{subequations}
    \begin{align}
        \Upsilon_{ijk}  &= \left\langle {a_i a_j a_k } \right\rangle_T \\
        \lambda_{ij}  &= \left\langle {a_i a_j } \right\rangle_T \\
        \mu_i  &= \left\langle {a_i } \right\rangle_T
    \end{align}
\end{subequations}
A positive rate of kinetic energy production, $\left\langle {\frac{d}{{dt}}e^{[1, \ldots ,n]}(t)} \right\rangle_T>0$ is associated with basis functions that resolve a greater proportion of large, energy-producing scales while a negative rate of kinetic energy production, $\left\langle {\frac{d}{{dt}}e^{[1, \ldots ,n]}(t)} \right\rangle_T<0$ is associated with basis functions that resolve a greater proportion of small, energy-dissipative scales. When the basis functions are POD basis functions, the rate of kinetic energy production tends to be positive, i.e. $\left\langle {\frac{d}{{dt}}e^{[1, \ldots ,n]}(t)} \right\rangle _T>0$. This is to be expected because the POD basis functions are biased toward the large, energy-producing scales of a turbulent flow. Galerkin-based ROMs of the Navier-Stokes equation deriving using only the first few most energetic POD basis functions tend to over predict the kinetic energy of the flow. The new proposed stabilization methodology achieves improved accuracy by deriving a new set of basis functions, ${\tilde a_i}$ ${\tilde {\bm{u}}_i}$ that model a greater proportion of small, energy-dissipation scales of the turbulent flow. The proposed stabilization methodology can be summarized as the following minimization problem
\begin{equation}
\label{eqn:full min problem}
    \begin{aligned}
    & \underset{{\tilde a},{\tilde {\bm{u}}}} {\operatorname{arg\,min}}
    & & \displaystyle \left\langle {\int\limits_\Omega  {\left| {\bm{u}_s (\bm{x},t) - \sum\limits_{i = 1}^n {{\tilde a}_i(t) {\tilde {\bm{u}}}_i(\bm{x})} } \right|^2 dx} } \right\rangle _T  \\
    & \text{s.t.}
    & & (1) \left\langle {{\tilde {\bm{u}}_i} ,{\tilde {\bm{u}}_j} } \right\rangle _\Omega = \delta _{ij} \\
    &&& (2) \displaystyle \left\langle {\frac{d}{{dt}}e^{[1, \ldots ,n]}(t) } \right\rangle_T = \epsilon
    \end{aligned}
\end{equation}
where, now, the turbulent kinetic energy, $e^{[1, \ldots ,n]}(t)$ is evaluated using the new basis functions, ${\tilde a_i}$ ${\tilde {\bm{u}}_i}$. The specific choice of the parameter $\epsilon$, referred to as the global dissipation parameter, is discussed in section~\ref{sec:Global dissipation parameter}. Since the objective function is infinitely dimensional, problem ~\eqref{eqn:full min problem} appears to be intractable. However, the infinitely dimensional objective function can be reduced to a finite dimensional objective function by assuming the $n$ new basis functions are spanned by the $N$ POD basis functions where $N>n$. In the spirit of related previous work~\citep{Amsallem:2012}, one can write
\begin{subequations}
    \begin{align}
        {\tilde U} &= UX \\
        {\tilde A} &= X^{T} A
  \end{align}
\end{subequations}
where $X \in \mathbb{R}^{N \times n}$ is an orthonormal ($X^{T} X = I_{n \times n}$) transformation matrix and the basis functions are vectorized and assembled in a matrix as follows
\begin{subequations}
  \begin{align}
      U  &= \left[ {\begin{array}{*{20}c} {\bm{u}_1 } & {\bm{u}_2 } &  \cdots  & {\bm{u}_N } \end{array}} \right] \\
      {\tilde U} &= \left[ {\begin{array}{*{20}c} {\tilde {\bm{u}}}_1 & {\tilde {\bm{u}}}_2  &  \cdots  & {\tilde {\bm{u}}}_n  \end{array}} \right] \\
      A  &= \left[ {\begin{array}{*{20}c} {a_1 } & {a_2 } &  \cdots  & {a_N }  \end{array}} \right]^T \\
      {\tilde A} &= \left[ {\begin{array}{*{20}c} {\tilde a}_1 & {\tilde a}_2 &  \cdots  & {\tilde a}_n  \end{array}} \right]^T
  \end{align}
\end{subequations}
Hence, the minimization problem \eqref{eqn:full min problem} can be formulated as the following finite dimensional problem
\begin{equation}
\label{eqn:full min problem2}
    \begin{aligned}
    & \underset{X \in \mathbb{R}^{N \times n}} {\operatorname{arg\,min}}
    & & \displaystyle \left\langle {\int\limits_\Omega  {\left| {\bm{u}_s (x,t) - U X X^{T} A } \right|^2 dx} } \right\rangle _T  \\
    & \text{s.t.}
    & & (1) X^T X = I_{n \times n} \\
    &&& (2) \displaystyle \left\langle {\frac{d}{{dt}}e^{[1, \ldots ,n]}(t) } \right\rangle_T = \epsilon
    \end{aligned}
\end{equation}
where the turbulent kinetic energy balance (i.e. constraint no. 2) is expressed as follows
\begin{equation}
    \left\langle {\frac{d}{{dt}}e^{[1, \ldots ,n]}(t) } \right\rangle _T  = \sum\limits_{i,j,k = 1}^n {{\tilde Q}_{ijk} {\tilde \Upsilon}_{ijk} }  + \sum\limits_{i,j = 1}^n {{\tilde L}_{ij} {\tilde \lambda}_{ij}  + } \sum\limits_{i = 1}^n {{\tilde b}_i {\tilde \mu}_i }
\end{equation}
The matrices ${\tilde Q}\in \mathbb{R}^{n \times n \times n}$, ${\tilde L} \in \mathbb{R}^{n \times n}$ and ${\tilde b} \in \mathbb{R}^{n}$ are the Galerkin matrices associated with the new spatial basis functions, ${\tilde {\bm{u}}}_i$ for $i=1,\cdots,n$, and ${\tilde \Upsilon} \in \mathbb{R}^{n \times n \times n}$, $\tilde {\lambda} \in \mathbb{R}^{n \times n}$ and ${\tilde \mu} \in \mathbb{R}^{n}$ are matrices containing the new temporal correlations
\begin{subequations}
    \begin{align}
        {\tilde \Upsilon}_{ijk}  &= \left\langle {{\tilde a}_i {\tilde a}_j {\tilde a}_k } \right\rangle_T \\
        {\tilde \lambda}_{ij}  &= \left\langle {{\tilde a}_i {\tilde a}_j } \right\rangle_T \\
        {\tilde \mu}_i  &= \left\langle {{\tilde a}_i } \right\rangle_T
    \end{align}
\end{subequations}
These new matrices can be all expressed as a function of the transformation matrix, $X$ as follows
\begin{subequations}
    \begin{align}
        {\tilde Q}_{ijk} &= \sum\limits_{p = 1}^N {X_{pi} X_{:i}^T Q_{p::} X_{:k} } \hspace{5mm} i,j,k=1,\cdots,n  \\
        {\tilde L} &= X^T L X \\
        {\tilde b} &= X^T b \\
        {\tilde \Upsilon}_{ijk} &= \sum\limits_{p = 1}^N {X_{pi} X_{:i}^T \Upsilon_{p::} X_{:k} } \hspace{5mm} i,j,k=1,\cdots,n \\
        {\tilde \lambda} &= X^T \lambda X \\
        {\tilde \mu} &= X^T \mu
    \end{align}
\end{subequations}
where $Q \in \mathbb{R}^{N \times N \times N}$, $L \in \mathbb{R}^{N \times N}$ and $b \in \mathbb{R}^{N}$ are the Galerkin matrices associated with the POD spatial basis functions, $\bm{u}_i$ for $i=1,\cdots,N$. \\
\indent To the best of the authors' knowledge, problem \eqref{eqn:full min problem2} does not admit any known closed form solutions and so, must be approximated numerically; details are provided in section~\ref{sec:The algorithm}. Furthermore, in its current formulation, solving problem \eqref{eqn:full min problem2} numerically would be computationally prohibitive and so further reductions in complexity are required. First, the objective function is reformulated. The objective function in problem \eqref{eqn:full min problem2} measures the error between the snapshot solution, $\bm{u}_s(\bm{x},t)$ and the spectral reconstruction, $\sum\nolimits_{i = 1}^n {{\tilde a}_i(t) {\tilde {\bm{u}}}_i(x)}$ or, in matrix notation, ${\tilde U\tilde A} = UXX^{T}A$. By definition, the reconstruction of the snapshot matrix using POD basis functions, $\sum\nolimits_{i = 1}^n {{a}_i(t) {{\bm{u}}}_i(x)}$ or, in matrix notation, $UA$ is the optimal reconstruction. Where by optimal it is meant that no other combination of the temporal and spatial basis functions provide a greater resolution of time-averaged, turbulent kinetic energy of the snapshot matrix. In other words, the following inequalities always hold
\begin{equation}
\left\langle {\int_\Omega  {\left| {\bm{u}_s (\bm{x},t)} \right|^2 dx} } \right\rangle _T  \ge \left\langle {\int_\Omega  {\left| {UA} \right|^2 dx} } \right\rangle _T  \ge \left\langle {\int_\Omega  {\left| {\tilde U\tilde A} \right|^2 dx} } \right\rangle _T
\end{equation}
Problem \eqref{eqn:full min problem2} can, therefore, be reformulated as follows
\begin{equation}
\label{eqn:full min problem3}
    \begin{aligned}
    & \underset{X \in \mathbb{R}^{N \times n}} {\operatorname{arg\,min}}
    & & \displaystyle \sum\limits_{i = 1}^N {\lambda_{ii} }  - \sum\limits_{i = 1}^n {\left( {X^T \lambda X} \right)_{ii} }  \\
    & \text{s.t.}
    & & (1) X^T X = I_{n \times n} \\
    &&& (2) \displaystyle \left\langle {\frac{d}{{dt}}e^{[1, \ldots ,n]}(t) } \right\rangle_T = \epsilon
    \end{aligned}
\end{equation}
where
\begin{subequations}
    \begin{align}
        \left\langle {\int_\Omega  {\left| {UA} \right|^2 dx} } \right\rangle _T  &= \sum\limits_{i = 1}^N {\lambda _{ii} }   \\
        \left\langle {\int_\Omega  {\left| {\tilde U\tilde A} \right|^2 dx} } \right\rangle _T  &= \sum\limits_{i = 1}^n {\tilde \lambda_{ii} }  = \sum\limits_{i = 1}^n {\left( {X^T \lambda X} \right)_{ii} }
    \end{align}
\end{subequations}
At this point, the second constraint of the minimization problem is reformulated. The most expensive calculation in the second constraint is the evaluation of the nonlinear Galerkin tensors, ${\tilde Q}\in \mathbb{R}^{n \times n \times n}$ and $Q\in \mathbb{R}^{N \times N \times N}$ which model the convective nonlinearity of the Navier-Stokes equations. As in any fully spectral discretization, approximation and evaluation of nonlinearities can be computationally prohibitive. This is particularly true in the proposed approach since the tensor, ${\tilde Q}$ must be evaluated at each iteration. Over the years, various approximations to tensor products have been derived~\citep{Carlberg:2011,Wang:2008}. In this paper, an alternative method to alleviate tensor computations is presented. It can be shown that for a large variety of flow boundary conditions, the Galerkin tensor, $Q$ is energy preserving, i.e. $\sum\nolimits_{ijk}^N {Q_{ijk} \left\langle {a_i a_j a_k } \right\rangle } _T  = 0$~\citep{Kraichnan:1989,Cordier:2012,Noack:2011}. For flows where this condition holds, it can be shown that transformed Galerkin tensor ${\tilde Q}$ is also energy preserving because a linear superposition of basis functions do not modify the boundary conditions. This nonlinear energy conversation principle holds for many boundary conditions including steady Dirichlet conditions, ambient flow or free-stream conditions at infinity, or periodic boundary conditions. For convective boundary condition, this energy conservation principle has not been rigorously proven. Fortunately, it has been observed to hold in a variety of numerical simulations of flows with convective boundary conditions. Hence, for these boundary conditions, problem \eqref{eqn:full min problem2} can be reformulated as follows
\begin{equation}
\label{eqn:full min problem4}
    \begin{aligned}
    & \underset{X \in \mathbb{R}^{N \times n}} {\operatorname{arg\,min}}
    & & \displaystyle \sum\limits_{i = 1}^N {\lambda _{ii} }  - \sum\limits_{i = 1}^n {\left( {X^T \lambda X} \right)_{ii} }  \\
    & \text{s.t.}
    & & (1) X^T X = I_{n \times n} \\
    &&& (2) \displaystyle \sum\limits_{i,j=1}^n {\left( X^T L X \circ X^T \lambda X \right)_{ij} }  = \epsilon
    \end{aligned}
\end{equation}
where $\circ$ is the Hadamard product. This final form of the minimization problem can be solved efficiently using standard numerical minimization algorithms.
\subsubsection{Global dissipation parameter, $\epsilon$}
\label{sec:Global dissipation parameter}
For a statistically stationary turbulent flow, the rate of turbulent kinetic energy production is, on average, zero, i.e. $\left\langle {\frac{d}{{dt}}e } \right\rangle _T = 0$. It would seem plausible, therefore, that an accurate ROM could be derived using a set of basis functions, ${\tilde a}$ and ${\tilde {\bm{u}}}$ for which $\left\langle {\frac{d}{{dt}}e^{[1, \ldots ,n]}(t) } \right\rangle _T = 0$ . Unfortunately, numerical simulations demonstrate that this not the case. Simulations suggest that ROM stability can be achieved using dissipative basis functions, i.e. basis functions for which $\left\langle {\frac{d}{{dt}}e^{[1, \ldots ,n]}(t) } \right\rangle _T = \epsilon$ where $\epsilon$ is a small negative scalar. In this paper, ROM stability is measured using the statistics of the turbulent kinetic energy. Specifically, ROM stability is quantified using the first moment of the turbulent kinetic energy as follows
\begin{equation}
\label{eqn:ROM stability}
    \left| {\sum\limits_{i = 1}^n {\tilde \lambda _{ii} }  - \left\langle {\sum\limits_{i = 1}^n {\tilde a_i }(t)^2 } \right\rangle _T } \right| = \delta _{\text{ROM}}
\end{equation}
where the temporal coefficients, ${\tilde a_i }(t)$ in \eqref{eqn:ROM stability} are derived using numerical integration of the ROM and $\delta _{\text{ROM}}$ is the convergence criteria. Numerical simulations suggest that ROM stability is a smooth function of the dissipation parameter, $\epsilon$ and therefore, a standard root-finding algorithm can be implemented. For both test cases summarized in this paper, convergence is achieved in less than $20$ iterations that, depending on the size of the ROM, equates to several minutes of CPU time.
\subsubsection{The algorithm and numerical implementation}
\label{sec:The algorithm}
For reasonably small problems, Eq.~\eqref{eqn:full min problem4} can be directly implemented and solved in MATLAB using Sequential Quadratic Programming (SQP)~\citep{Dixon:1975,Schittkowski:1986,Nocedal:1999} and Brent's methods~\citep{Brent:2002}; MATLAB functions \verb=fmincon= and \verb=fzero= respectively. Details of the MATLAB implementation are summarized in the appendix. The following pseudo-code summarizes the approach.
\begin{algorithm}[h!tb]
    \caption{Stabilization Algorithm\label{Alg:stabilization new criteria}}
    \begin{algorithmic}[1]
        \REQUIRE POD basis functions, $\bm{u}_i(\bm{x})$ and $a_i(t)$ for $i=1,\cdots,N$ and the associated Galerkin matrices $Q \in \mathbb{R}^{N \times N \times N}$, $L \in \mathbb{R}^{N \times N}$ and $b \in \mathbb{R}^{N}$. The global dissipation parameter $\epsilon$,
        \ENSURE Galerkin matrices ${\tilde Q}\in \mathbb{R}^{n \times n \times n}$, ${\tilde L} \in \mathbb{R}^{n \times n}$ and ${\tilde b} \in \mathbb{R}^{n}$ associated with the transformed basis functions, ${\tilde {\bm{u}}}_i$ for $i = 1, \cdots, n$
        \STATE Choose $n>0$, s.t. $N>n$
        \STATE Compute the POD eigenvalues, $\lambda _{ij}  = \left\langle {a_i a_j } \right\rangle _T$
        \WHILE{\text{ROM is unstable, i.e.}$\left| {\sum\nolimits_{i = 1}^n {\tilde \lambda _{ii} }  - \left\langle {\sum\nolimits_{i = 1}^n {\tilde a_i }(t)^2 } \right\rangle _T } \right| > \delta _{\text{ROM}} $}
            \STATE Solve the optimization problem:
                    \begin{equation*}
                        \begin{aligned}
                        & \underset{X \in \mathbb{R}^{N \times n}} {\operatorname{arg\,min}}
                        & & \displaystyle \sum\limits_{i = 1}^N {\lambda _i }  - \sum\limits_{i = 1}^n {\left( {X^T \lambda X} \right)_{ii} }  \\
                        & \text{s.t.}
                        & & (1) X^T X = I_{n \times n} \\
                        &&& (2) \displaystyle \sum\limits_{i,j=1}^n {\left( X^T L X \circ X^T \lambda X \right)_{ij} }  = \epsilon
                        \end{aligned}
                    \end{equation*}
            \STATE Evaluate new Galerkin matrices, ${\tilde Q}$, ${\tilde L}$ and ${\tilde b}$.
            \STATE Integrate ROM, check for stability.
            \STATE Update $\epsilon$ using Brent's method.
        \ENDWHILE
        \STATE Return $\bm{u}'_i$, $X$ and new Galerkin matrices ${\tilde Q}$, ${\tilde L}$ and ${\tilde b}$
    \end{algorithmic}
\end{algorithm}
\section{Applications}
\label{chap:Applications}
\subsection{Lid-driven cavity}
\label{sec:Lid-driven cavity_data}
\begin{figure}
\centering
    \begin{tikzpicture}
        \begin{axis}[
        ybar, xbar,
            axis on top,
            enlargelimits=false,
            width=6cm,
            height=6cm,
            xlabel={$x$},
            ylabel={$y$},
            ]
                \addplot graphics
                [xmin=-1,xmax=1,ymin=-1,ymax=1,
                includegraphics={trim=0 0 0 0,clip},]
                {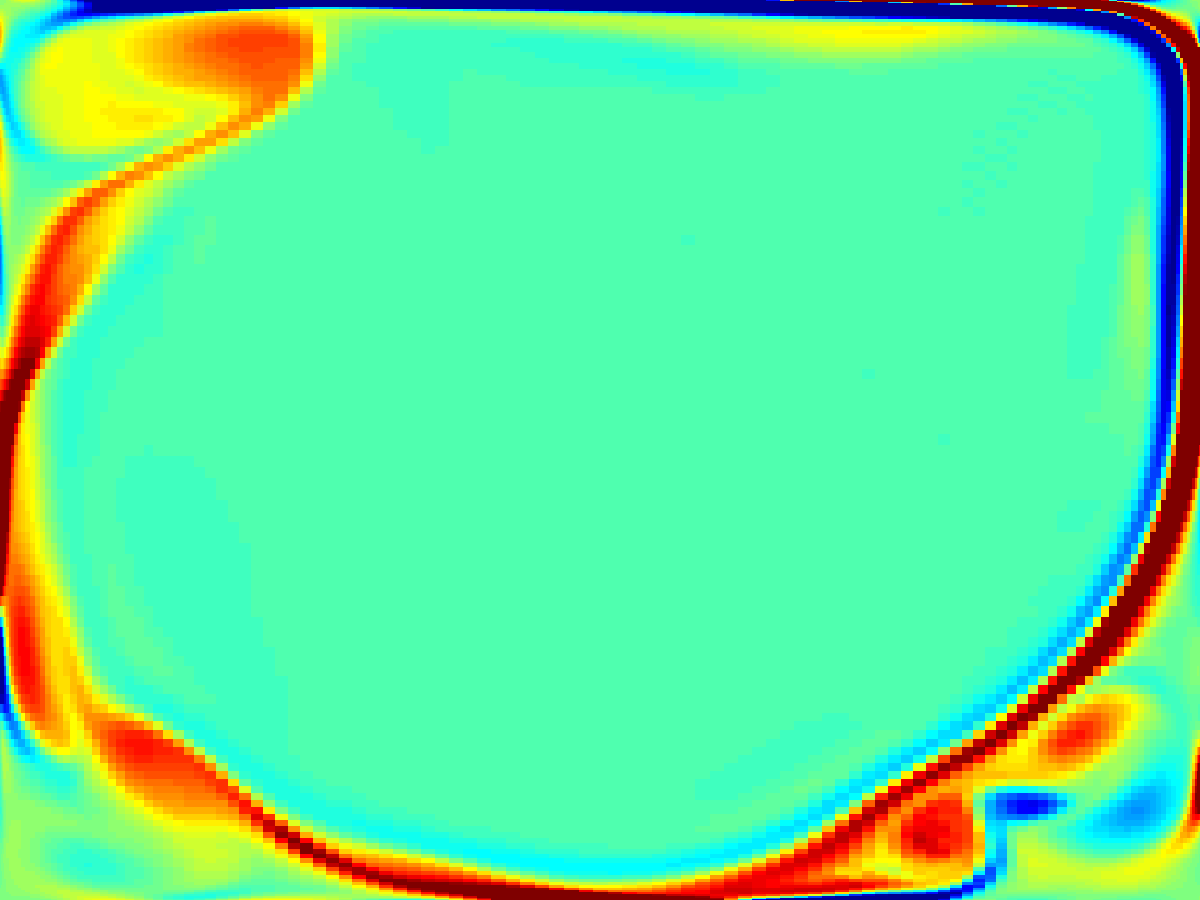};
        \end{axis}
    \end{tikzpicture}
\caption{Vorticity contours from a DNS of the lid-driven cavity at $\mathop{\rm Re_u}=3\times10^4$\label{fig:lid_snapshot}}
\end{figure}

The lid-driven cavity is a well known benchmark problem used to validate fluid flow numerical schemes and reduced order models~\citep{Cazemier:1998,Shankar:2000,Terragni:2011}. Specifically, isothermal, incompressible, two-dimensional flow inside a square cavity driven by a prescribed lid velocity, $\bm{u}_{lid}=(1-x^2)^2$ is considered. The Reynolds number (Re) is defined with respect to the maximum velocity of the lid and the width of the cavity. The Navier-Stokes are discretized in space using Chebyshev polynomials. The convective nonlinearities are handled pseudo-spectrally and the Chebyshev coefficients are derived using the Fast Fourier Transform (FFT). The Navier-Stokes are discretized in time using a semi-implicit, second-order Euler scheme. Figure~\ref{fig:lid_snapshot} is a snapshot of a statistically stationary solution at $\mathop{\rm Re_{u}}=3\times10^4$. This particular simulation was performed using a $128^2$ Chebyshev grid. The computations were performed on a cluster at Duke University using $8$ processors. The simulation is first initialized over $100\,000$ time steps ($\Delta t=1 \times 10^{-3}\text{s}$) which corresponds to a total run time of $1$ hours. The data base is then generated: $50\,000$ iterations are done on $0.5$ hours to create $5000$ snapshots.\\
\subsubsection{POD basis functions of the lid-driven cavity}
\label{sec:Lid-driven cavity modes}
\begin{figure}
\centering
        \subfigure[$\bm{u}_1$\label{sufig:lid_w_1}]
            {
            \begin{tikzpicture}
                \begin{axis}[
                ybar, xbar,
                    axis on top,
                    enlargelimits=false,
                    width=4cm,
                    height=4cm,
                    xlabel={$x$},
                    ylabel={$y$},
                    ]
                        \addplot graphics
                        [xmin=-1,xmax=1,ymin=-1,ymax=1,
                        includegraphics={trim=0 0 0 0,clip},]
                        {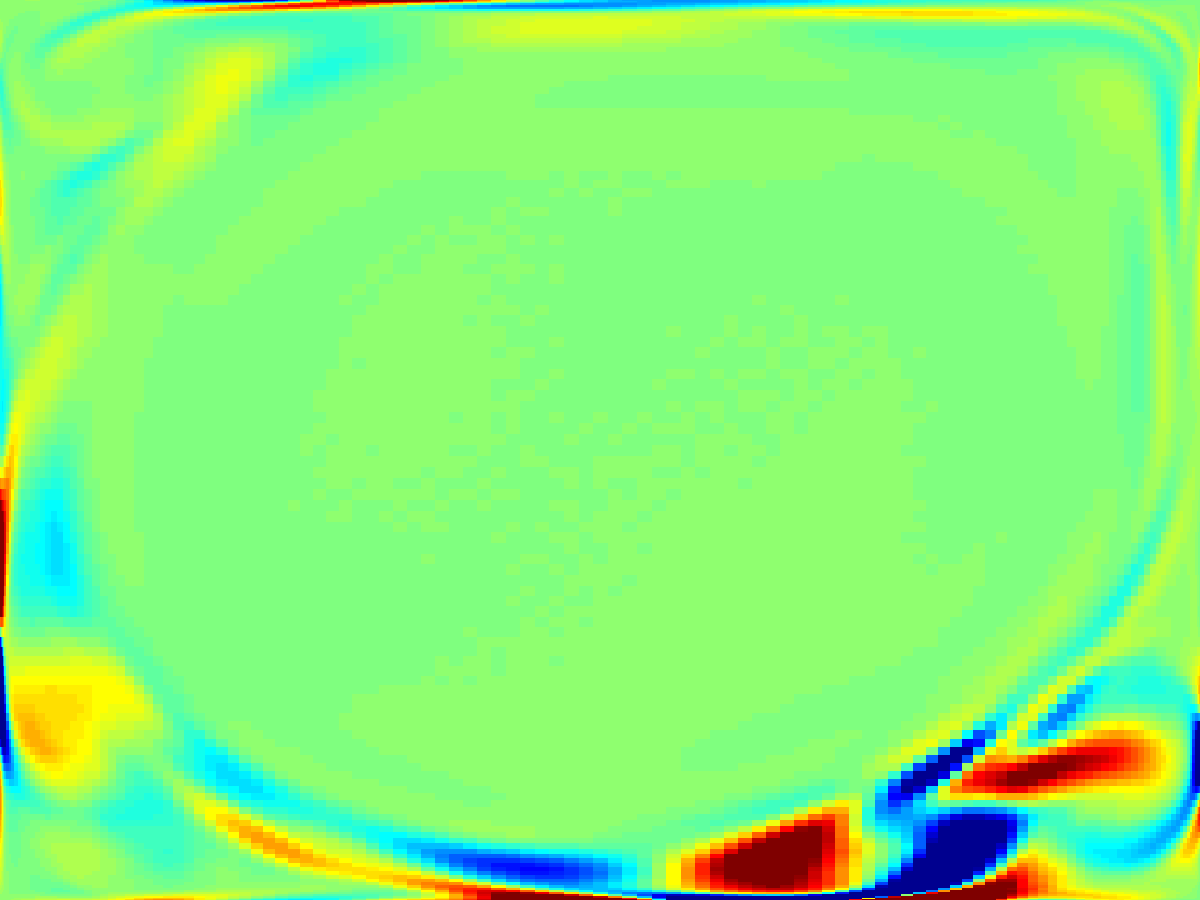};
                \end{axis}
            \end{tikzpicture}
            }
        \subfigure[$\bm{u}_2$\label{sufig:lid_w_2}]
            {
            \begin{tikzpicture}
                \begin{axis}[
                ybar, xbar,
                    axis on top,
                    enlargelimits=false,
                    width=4cm,
                    height=4cm,
                    xlabel={$x$},
                    ylabel={$y$},
                    ]
                        \addplot graphics
                        [xmin=-1,xmax=1,ymin=-1,ymax=1,
                        includegraphics={trim=0 0 0 0,clip},]
                        {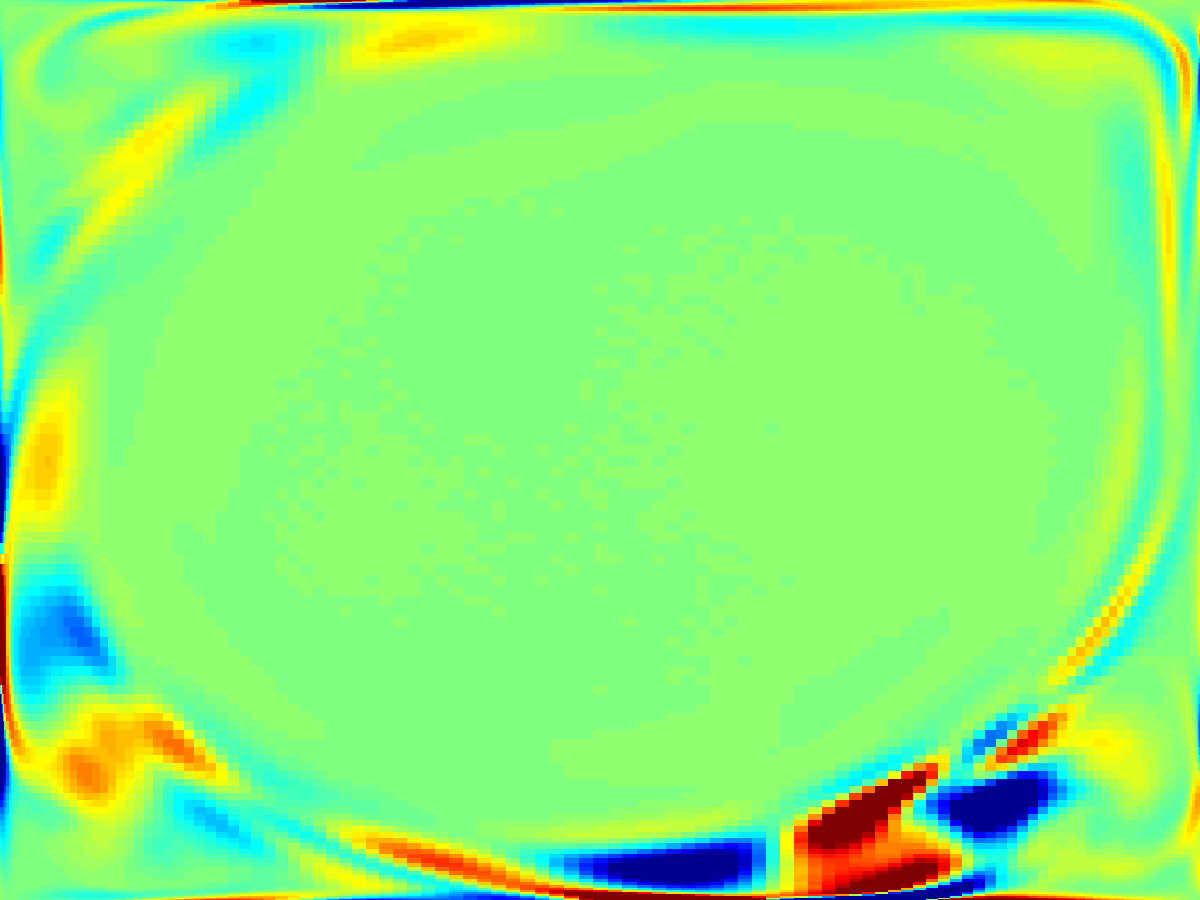};
                \end{axis}
            \end{tikzpicture}
        }
        \subfigure[$\bm{u}_{10}$\label{sufig:lid_w_10}]
            {
            \begin{tikzpicture}
                \begin{axis}[
                ybar, xbar,
                    axis on top,
                    enlargelimits=false,
                    width=4cm,
                    height=4cm,
                    xlabel={$x$},
                    ylabel={$y$},
                    ]
                        \addplot graphics
                        [xmin=-1,xmax=1,ymin=-1,ymax=1,
                        includegraphics={trim=0 0 0 0,clip},]
                        {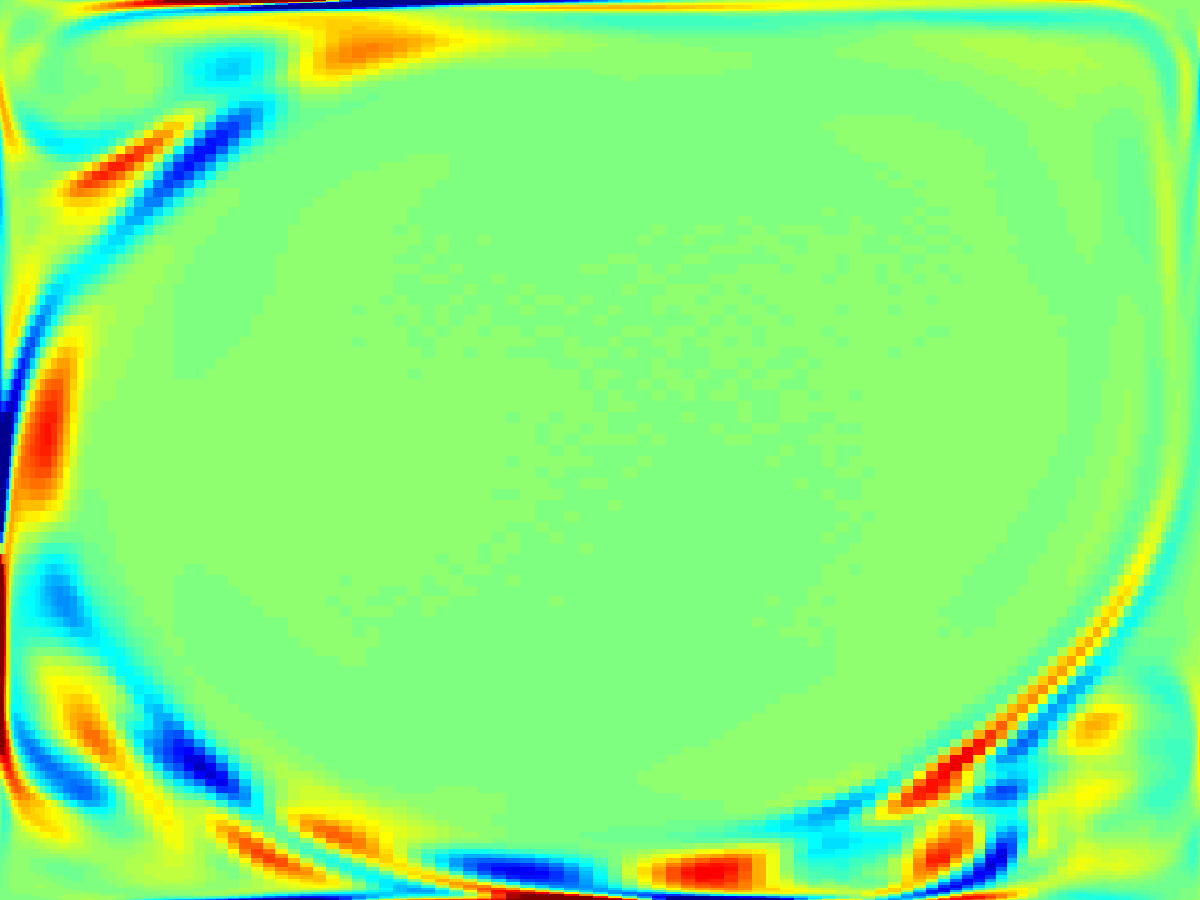};
                \end{axis}
            \end{tikzpicture}
        }
        \subfigure[$\bm{u}_{20}$\label{sufig:lid_w_20}]
            {
            \begin{tikzpicture}
                \begin{axis}[
                ybar, xbar,
                    axis on top,
                    enlargelimits=false,
                    width=4cm,
                    height=4cm,
                    xlabel={$x$},
                    ylabel={$y$},
                    ]
                        \addplot graphics
                        [xmin=-1,xmax=1,ymin=-1,ymax=1,
                        includegraphics={trim=0 0 0 0,clip},]
                        {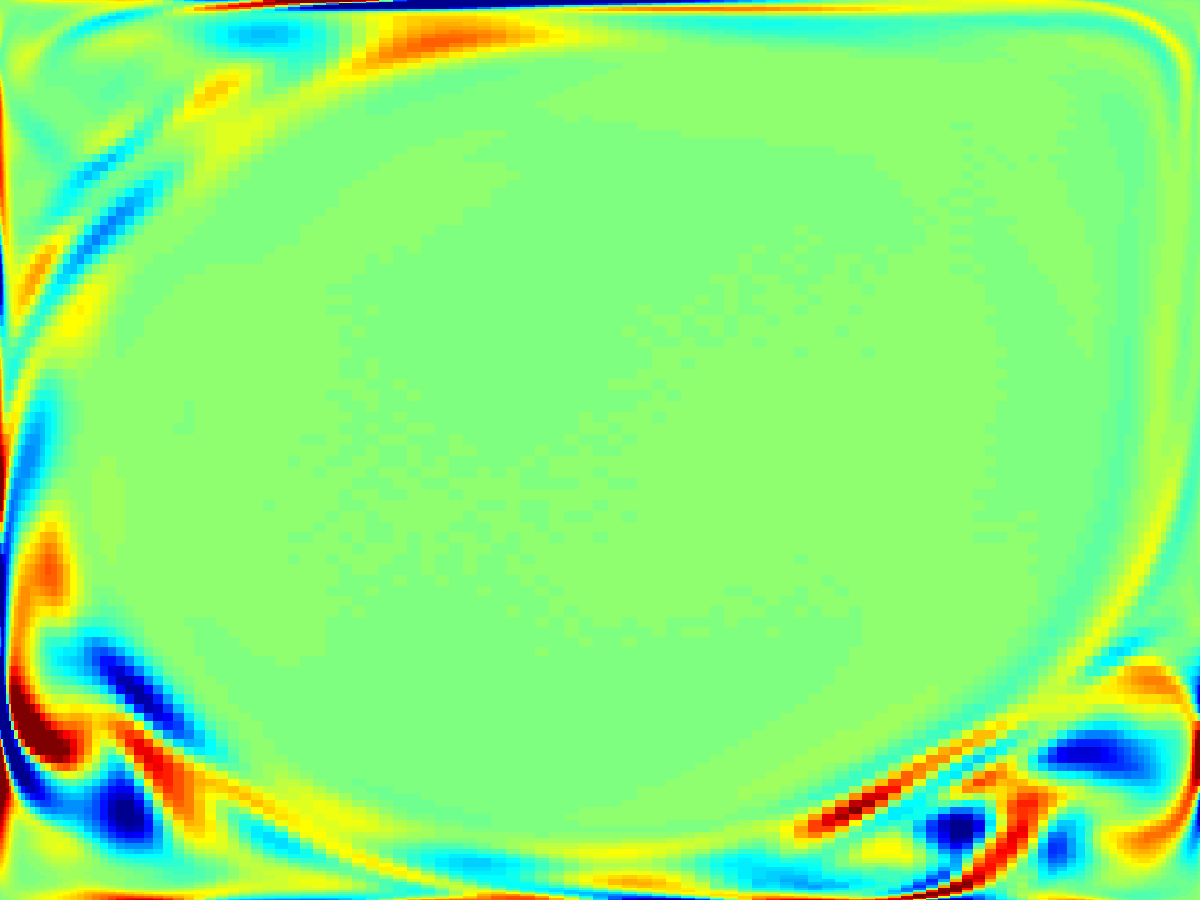};
                \end{axis}
            \end{tikzpicture}
        }
        \subfigure[$\bm{u}_{50}$\label{sufig:lid_w_50}]
            {
            \begin{tikzpicture}
                \begin{axis}[
                ybar, xbar,
                    axis on top,
                    enlargelimits=false,
                    width=4cm,
                    height=4cm,
                    xlabel={$x$},
                    ylabel={$y$},
                    ]
                        \addplot graphics
                        [xmin=-1,xmax=1,ymin=-1,ymax=1,
                        includegraphics={trim=0 0 0 0,clip},]
                        {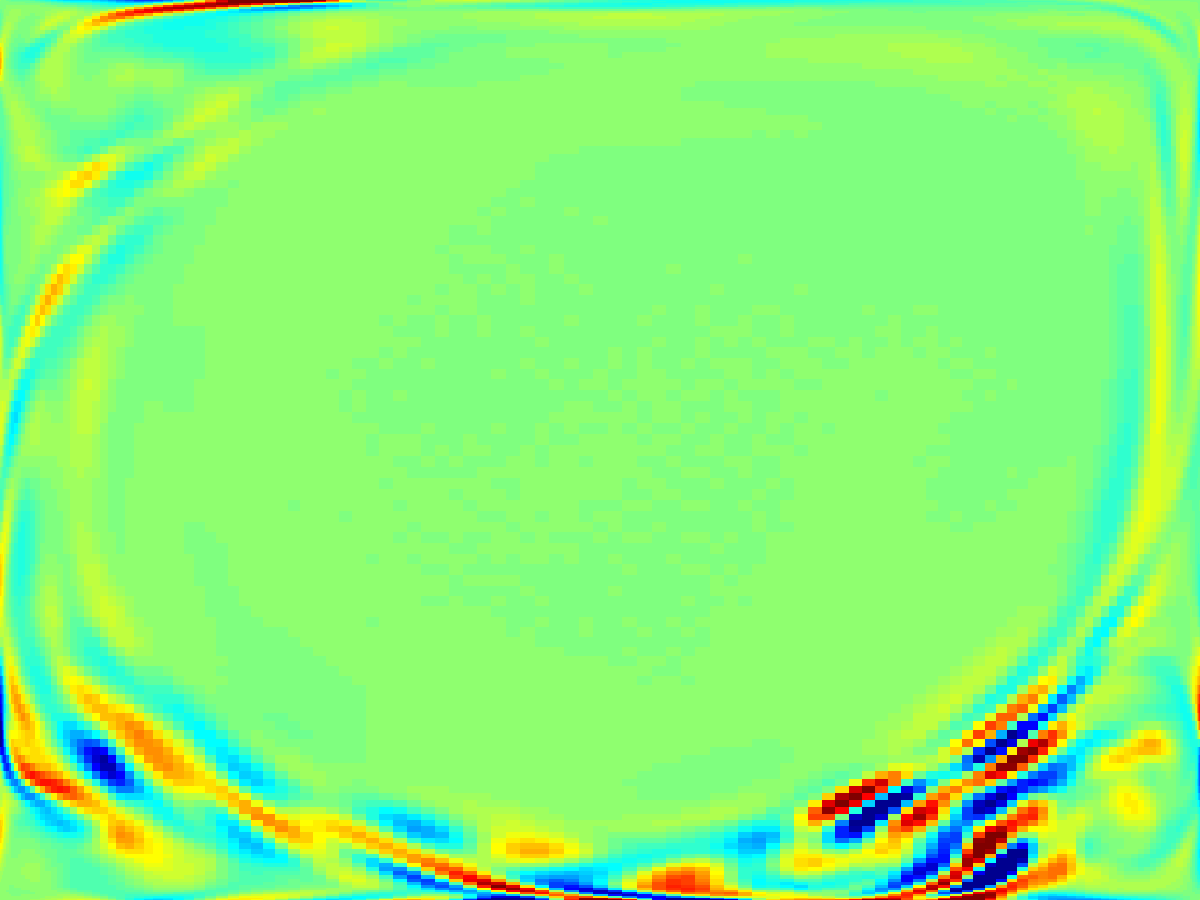};
                \end{axis}
            \end{tikzpicture}
        }
        \subfigure[$\bm{u}_{200}$\label{sufig:lid_w_250}]
            {
            \begin{tikzpicture}
                \begin{axis}[
                ybar, xbar,
                    axis on top,
                    enlargelimits=false,
                    width=4cm,
                    height=4cm,
                    xlabel={$x$},
                    ylabel={$y$},
                    ]
                        \addplot graphics
                        [xmin=-1,xmax=1,ymin=-1,ymax=1,
                        includegraphics={trim=0 0 0 0,clip},]
                        {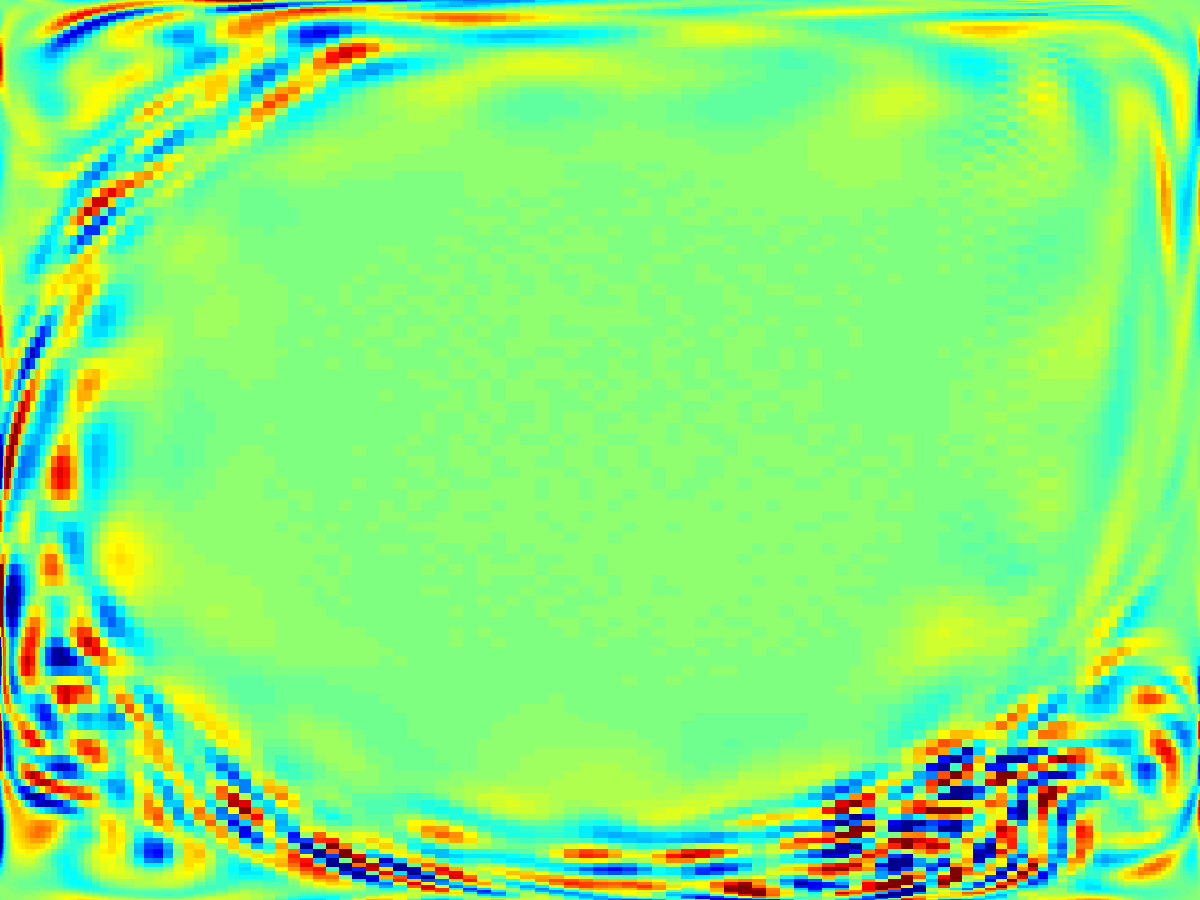};
                \end{axis}
            \end{tikzpicture}
        }
\caption{Vorticity contours of spatial POD basis functions of the lid-driven cavity at $\mathop{\rm Re_u}=3\times10^4$.} \label{fig:lid_POD_spatial_basis_functions}
\end{figure}

A database of 5000 DNS snapshots of the lid-driven cavity were used to find the POD basis functions, $\bm{u}_i(\bm{x})$ and $a_i(t)$. The nondimensional time interval between each snapshot equaled 0.1. Increasing the number of snapshots or the time interval between each snapshot had no significant effect on the performance characteristics of the ROMs. The percentaged of time-averaged, turbulent kinetic energy captured by the first $n$ POD basis functions is labeled, $E^{[1, \ldots ,n]}$ and defined as follows
\begin{equation}
    \displaystyle E^{[1, \ldots ,n]}  = \frac{{\left\langle {e^{[1, \ldots ,n]} } \right\rangle _T }}{{\left\langle {e_s } \right\rangle _T }} \times 100 = \displaystyle \frac{{\displaystyle \frac{1}{2}\displaystyle \int\limits_\Omega  {\left| {\displaystyle \sum\limits_{i = 1}^n {a(t)\bm{u}(\bm{x})} } \right|^2 dx} }}{{\displaystyle \frac{1}{2} \displaystyle \int\limits_\Omega  {\left| {\bm{u}_s (\bm{x},t)} \right|^2 dx} }} \times 100 = \displaystyle \frac{{\displaystyle \sum\limits_{i = 1}^n {\lambda _{ii} } }}{{\displaystyle \sum\limits_{i = 1}^{\infty   } {\lambda _{ii} } }} \times 100
\end{equation}
Results for the lid-driven cavity are shown in Table~\ref{tab:lid_eigenvalues}.
\begin{table}
    \begin{center}
    \def~{\hphantom{0}}
        \begin{tabular}{c c}
            n & $E^{[1, \ldots ,n]}$ \\
            1 & 16.06 \\
            2 & 29.21 \\
            3 & 37.45 \\
            4 & 44.88 \\
            5 & 50.37 \\
            10 & 67.16 \\
            20 & 82.40 \\
            50 & 93.21 \\
            200 & 99.31 \\
        \end{tabular}
    \end{center}
    \caption{Percent of time-averaged, turbulent kinetic energy captured by the first $n$ POD basis functions of the lid-driven cavity at $\mathop{\rm Re_u}=3\times10^4$.}\label{tab:lid_eigenvalues}
\end{table}
Vorticity contours of spatial POD basis functions, $\bm{u}_i(\bm{x})$ for $i=1,2,20,50$ and $200$ of the lid-driven cavity are illustrated in Fig.~\ref{fig:lid_POD_spatial_basis_functions}. As expected, the overall spatial wavelengths of the POD basis functions tends to increases with order. The low-order POD basis functions correspond to the large, high-energy physical scales of the turbulent flows while the higher-order POD basis functions correspond to the small, low-energy physical scales of the turbulent flow.
\subsubsection{Galerkin ROMs of lid-driven cavity using standard POD basis functions, $a_i(t)$ and $\bm{u}_i(\bm{x})$}
\label{sec:Lid-driven cavity_standard_ROMS}
\begin{figure}
\centering
    \begin{tikzpicture}
        \begin{semilogyaxis}[
            ymin=0.0001,
            ymax=1e2,
            width=12cm,
            height=5cm,
            xlabel={Time},
            ylabel={ $e(t)$ },
            y label style={at={(-0.02,0.5)}},
            legend columns=1,
            ]
            \addplot+[line join=round][
                line width=2pt,
                color=gray,
                mark={false},
                ]
                file {data/LID_E_DNS.dat}; \label{legend:thick_gray}
            \addplot+[line join=round][
                line width=1pt,
                color=red,
                mark={false},
                ]
                file {data/LID_E_ROM_Standard_0005.dat}; \label{legend:thin_red}
            \addplot+[line join=round][
                line width=1pt,
                color=blue,
                mark={false},
                ]
                file {data/LID_E_ROM_Standard_0010.dat}; \label{legend:thin_blue}
            \addplot+[line join=round][
                line width=2pt,
                color=black,
                mark={false},
                dashed,
                ]
                file {data/LID_E_ROM_Standard_0050.dat}; \label{legend:dashed_black}
            \addplot+[line join=round][
                line width=1pt,
                color=black,
                mark={false},
                ]
                file {data/LID_E_ROM_Standard_0200.dat}; \label{legend:thin_black}
         \end{semilogyaxis}
    \end{tikzpicture}
\caption{Galerkin ROMs of the lid-driven cavity derived using the first $n=5$(\ref{legend:thin_red}), $10$(\ref{legend:thin_blue}), $50$(\ref{legend:dashed_black}) and $200$(\ref{legend:thin_black}) POD basis functions; DNS (\ref{legend:thick_gray})}\label{fig:LID_e_STANDARD}
\end{figure}
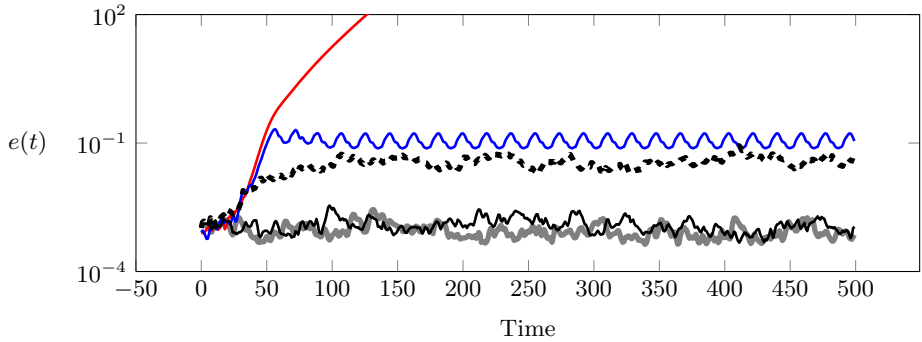

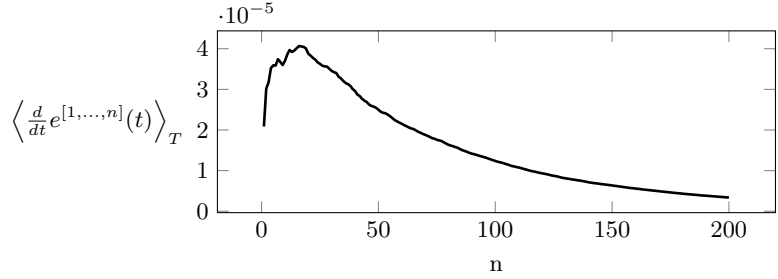
\begin{figure}
    \centering
            \begin{tikzpicture}
                \begin{axis}[
                    scaled ticks=true,
                    width=9cm,
                    height=4cm,
                    xlabel={n},
                    ylabel={ $\left\langle {\frac{d}{{dt}}e^{[1, \ldots ,n]}(t) } \right\rangle_T$ },
                    y label style={at={(-0.05,0.5)}}
                    ]
                    \addplot+[line join=round][
                        line width=1pt,
                        color=black,
                        mark={false},
                        ]
                        file {data/LID_rate_of_e.dat};
                 \end{axis}
            \end{tikzpicture}
    \caption{Time-averaged rate of turbulent kinetic energy production associated with the first $n$ POD basis functions of the lid-driven cavity}\label{fig:LID_rate_of_e}
\end{figure}

In this section, Galerkin ROMs of the lid-driven cavity are derived using the standard POD basis functions, $a_i(t)$ and $\bm{u}_i(\bm{x})$ for $i=1,\cdots,n$. The turbulent kinetic energy as predicted by these ROMs is illustrated in Figure~\ref{fig:LID_e_STANDARD}. Numerical integration of the ROM was performed in MATLAB using built-in adaptive forth/fifth order Runge-Kutta scheme \verb=ODE45=. Despite the fact that POD basis functions capture a large percentage of the time-averaged, turbulent kinetic energy of the snapshot solution (See Table~\ref{tab:lid_eigenvalues}), Figure~\ref{fig:LID_e_STANDARD} clearly indicates Galerkin ROMs based on these basis functions do not converge to the DNS. Convergence is achieved only when a prohibitively large number of POD basis functions are utilized; approximately $n=200$ POD basis functions for this particular test case.
These convergence issues, of course, were anticipated. Galerkin ROMs of the Navier-Stokes that utilize only the first few most energetic POD basis functions tend to under-resolve the small, energy-dissipating scales of the turbulent flow which, therefore, leads to excessive kinetic energy production. As discussed in Section~\ref{sec:Reduced Order Model (ROM) stability and accuracy}, the time-averaged rate of kinetic energy production associated with the POD basis functions can be calculated using Eq.~\eqref{Eqn:rate_of_e}; these rates are illustrated in Figure~\ref{fig:LID_rate_of_e}. As shown, the time-averaged rate of kinetic energy production associated with the first $n$ POD basis functions is always positive and approaches zero asymptotically. It is not surprising, therefore, that ROMs derived using only the most energetic basis functions tend to over predict the kinetic energy of the flow.
\subsubsection{Galerkin ROMs of lid-driven cavity using the new basis functions, ${\tilde a_i(t)}$ and ${\tilde {\bm{u}}_i(\bm{x})}$}
\label{sec:Lid-driven cavity_new_ROM}
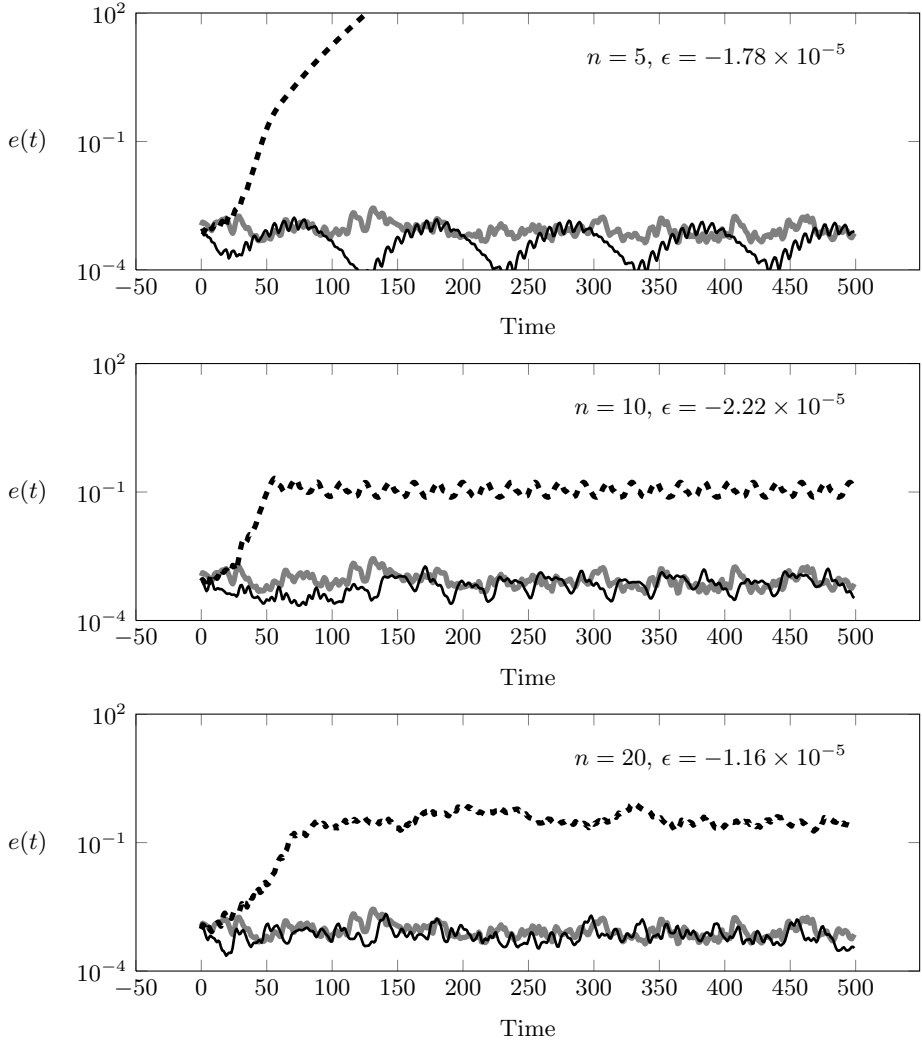
\begin{figure}
\centering
    \begin{tikzpicture}
        \begin{semilogyaxis}[
            ymin=0.0001,
            ymax=1e2,
            width=12cm,
            height=5cm,
            xlabel={Time},
            ylabel={ $e(t)$ },
            y label style={at={(-0.02,0.5)}},
            legend columns=1,
            ]
            \addplot+[line join=round][
                line width=2pt,
                color=gray,
                mark={false},
                ]
                file {data/LID_E_DNS.dat}; \label{legend:thick_gray}
            \addplot+[line join=round][
                line width=2pt,
                color=black,
                dashed,
                mark={false},
                ]
                file {data/LID_E_ROM_0005_0005_0.dat}; \label{legend:dashed_black}
            \addplot+[line join=round][
                line width=1pt,
                color=black,
                mark={false},
                ]
                file {data/LID_E_ROM_0005_0020_1.dat}; \label{legend:thin_black}
                \draw (axis cs:500,1e1) node[left]
                {$n = 5$, $\epsilon = -1.78\times 10^{-5}$};
         \end{semilogyaxis}
    \end{tikzpicture}
    \begin{tikzpicture}
        \begin{semilogyaxis}[
            ymin=0.0001,
            ymax=1e2,
            width=12cm,
            height=5cm,
            xlabel={Time},
            ylabel={ $e(t)$ },
            y label style={at={(-0.02,0.5)}},
            legend columns=1,
            ]
            \addplot+[line join=round][
                line width=2pt,
                color=gray,
                mark={false},
                ]
                file {data/LID_E_DNS.dat};
            \addplot+[line join=round][
                line width=2pt,
                color=black,
                dashed,
                mark={false},
                ]
                file {data/LID_E_ROM_0010_0010_0.dat};
            \addplot+[line join=round][
                line width=1pt,
                color=black,
                mark={false},
                ]
                file {data/LID_E_ROM_0010_0020_1.dat};
                \draw (axis cs:500,1e1) node[left]
                {$n = 10$, $\epsilon = -2.22\times 10^{-5}$};
         \end{semilogyaxis}
    \end{tikzpicture}
    \begin{tikzpicture}
        \begin{semilogyaxis}[
            ymin=0.0001,
            ymax=1e2,
            width=12cm,
            height=5cm,
            xlabel={Time},
            ylabel={ $e(t)$ },
            y label style={at={(-0.02,0.5)}},
            legend columns=1,
            ]
            \addplot+[line join=round][
                line width=2pt,
                color=gray,
                mark={false},
                ]
                file {data/LID_E_DNS.dat};
            \addplot+[line join=round][
                line width=2pt,
                color=black,
                dashed,
                mark={false},
                ]
                file {data/LID_E_ROM_0020_0020_0.dat};
            \addplot+[line join=round][
                line width=1pt,
                color=black,
                mark={false},
                ]
                file {data/LID_E_ROM_0020_0050_1.dat};
                \draw (axis cs:500,1e1) node[left]
                {$n = 20$, $\epsilon = -1.16\times 10^{-5}$};
         \end{semilogyaxis}
    \end{tikzpicture}
\caption{Evolution of the turbulent kinetic energy of the lid driven cavity as predicted by the DNS (\ref{legend:thick_gray}), a standard POD-based Galerkin ROM (\ref{legend:dashed_black}) and the new Galerkin ROM (\ref{legend:thin_black}).}\label{fig:LID_e}
\end{figure}

\indent In this section, ROMs of the lid-driven cavity are derived using the new proposed methodology. Galerkin ROMs are derived using basis functions with \textit{negative} rates of kinetic energy production, i.e. basis functions for which $\left\langle {\frac{d}{{dt}}e^{[1, \ldots ,n]}(t)} \right\rangle_T = \epsilon$ where $\epsilon$ is a negative scalar. In the previous section it was demonstrated that the standard POD basis functions have positive rates of kinetic energy production, $\left\langle {\frac{d}{{dt}}e^{[1, \ldots ,n]}(t)} \right\rangle_T>0$ and therefore, produce Galerkin ROMs that over predict the kinetic energy of the flow. The critical rate of kinetic energy production, $\epsilon$ for each ROM order $n$ is found iteratively using the algorithm introduced in Section~\ref{sec:The algorithm}.\\
\indent Figure~\ref{fig:LID_e} illustrates the evolution of the turbulent kinetic energy of lid-driven cavity as predicted by the new Galerkin ROMs. For the sake of brevity, only $n=5$, 10 and 20 ROMs are illustrated. The turbulent kinetic energy of the DNS solution is identified by the thick grey curves while the dashed and solid black curves identify the turbulent kinetic energy of the standard and new ROMs respectively. As predicted, ROMs derived using the new basis functions converge to the correct mean value of kinetic energy and the accuracy of the models is improved as $n$ is increased. As stated previously, these new basis functions have negative rates of kinetic energy production, $\epsilon$ while the standard POD basis functions have positive rates of kinetic energy production. Despite this difference, these new basis functions remain very similar to the POD basis functions as summarized in Table~\ref{tab:LID_e of modes}. The percentaged of time-averaged, turbulent kinetic energy captured by the first $n$ new basis functions is labeled, $\tilde E^{[1, \ldots ,n]}$ and defined as follows
\begin{equation}
\tilde E^{[1, \ldots ,n]}  = \frac{{\sum\limits_{i = 1}^n {\tilde \lambda _{ii} } }}{{\sum\limits_{i = 1}^{\infty  } {\lambda _{ii} } }} \times 100
\end{equation}
For all orders $n$, the new basis functions capture a very similar total of kinetic energy of the flow. In other words, the transformation matrix associated with these new basis functions resembles the following
\begin{equation}
    X_{N \times n}  = \left[ {\begin{array}{*{20}c}
       I_{n \times n}  \\
       O_{(N - n) \times n} \\
    \end{array}} \right] + \delta_{N \times n}
\end{equation}
where $I_{n \times n}$ and $O_{(N - n) \times n}$ are the identity and null matrices respectively. $\delta _{N \times n} $ is a matrix whose entries are all less than one, i.e. $\delta_{ij} < 1$ $\forall i,j$. Therefore, the new basis function inherent much of the optimality of the original POD basis function.
\begin{table}
    \begin{center}
    \def~{\hphantom{0}}
        \begin{tabular}{l c c}
            n & $E^{[1, \ldots ,n]}$ &  $\tilde E^{[1, \ldots ,n]}$ \\
            5 & 50.37 & 50.01 \\
            10 & 67.16 & 66.72\\
            20 & 82.40 & 82.11\\
        \end{tabular}
    \end{center}
    \caption{Percentage of time-averaged, turbulent kinetic energy captured by the first $n$ basis functions of the lid-driven cavity.}\label{tab:LID_e of modes}
\end{table}

\subsection{Mixing layer}
\label{sec:Mixing layer_data}
The data base used for the present work corresponds to the DNS of an isothermal 2-D mixing layer. The numerical algorithm is the same as that employed previously for studies on jet noise sources~\citep{Cavalieri:2011}. The full Navier-Stokes equations for 2-D fluid motion are formulated in Cartesian coordinates and solved in conservative form. Spatial derivatives are computed with a fourth-order-accurate finite scheme for both the inviscid and viscous portion of the flux~\citep{Hayder:1993,Gottlieb:1976}. A second-order predictor-corrector scheme is used to advance the solution in time. In addition, block decomposition and MPI parallelization are implemented. The three-dimensional Navier-Stokes characteristic non-reflective boundary conditions (3D-NSCBC), developed in~\citep{Lodato:2007}, are applied at the boundaries of the computational domain to account for convective fluxes and pressure gradients across the boundary plane. In order to simulate anechoic boundary conditions, the mesh is stretched and a dissipative term is added to the equations, in the sponge zone~\citep{Colonius:1993}. A detailed description of the numerical procedure is given in~\citep{Daviller:2010}.\\
\indent The inflow mean streamwise velocity profile is given by a hyperbolic tangent profile
\begin{equation}
  \overline{u}(y)=U_2+\frac{1}{2}\Delta U \left[ 1 + \tanh(2y) \right]
\end{equation}
with $\Delta U=U_1-U_2$ the velocity difference across the mixing layer, where $U_1$ and $U_2$ are the initial velocity above and below, respectively. The velocities, lengths, and time are nondimensionalized with $\Delta U$ and the initial vorticity thickness $\delta_{\omega}$. The flow Reynolds number is $Re=\delta_{\omega} \Delta U/ \nu_{a}=500$ where the subscript $\left(\cdot\right)_a$ indicates a constant ambient quantity.  The Mach number of the free streams are $M_1=U_1/c_{a}=0.1$ and $M_2=U_2/c_{a}=0.033$, where $c_{a}$ is the sound speed. The inflow mean temperature is calculated with the Crocco-Busemann relation, and the inflow mean pressure is constant. The Prandtl number is selected to be $Pr=0.7$. Finally, the convective Mach number is given by $M_c=\Delta U/2c_{a}=0.033$, so that the flow can be assumed as quasi-incompressible. The numerical code was extensively validated against numerical and experimental data; some results can be found in~\citep{Cavalieri:2011,Daviller:2010}.
The computations were performed on the cluster of the PPRIME Institute in Poitiers, France using $64$ processors. The computational domain comprises approximately $2.1$ million grid points: $2367$ points in the streamwise direction, and $884$ points along the $y$ direction. The extension of the computational domain
is $325 \delta_{\omega} \times 120 \delta_{\omega}$. The sponge regions are from $x=-20\delta_{\omega}$ to $x=0$ and $x=250\delta_{\omega}$ to $x=305\delta_{\omega}$ in the streamwise direction, and from $\pm 50\delta_{\omega}$ to $\pm 60\delta_{\omega}$ in the transverse $y$ direction. To promote a natural transition to turbulence from an initially laminar solution, the flow is forced by adding at every iteration solenoidal perturbations defined as \citep{Bogey:2000}.The simulation is first initialized over $330\,000$ time steps ($\Delta t=1.8 \times 10^{-8}\text{s}$) which corresponds to a total run time of $35$ hours. The data base is then generated: $1\,093\,695$ iterations are done on $115$ hours to create $2000$ snapshots.\\
\begin{figure}
\centering
    \begin{tikzpicture}
        \begin{axis}[
        ybar, xbar,
            axis on top,
            enlargelimits=false,
            width=12cm,
            height=5cm,
            xlabel={$x$},
            ylabel={$y$},
            xtick={0,50,100,150,200,250,300},
            ytick={-50,0,50},
            ]
                \addplot graphics
                [xmin=0,xmax=300,ymin=-50,ymax=50,
                includegraphics={trim=0 0 0 0,clip},]
                {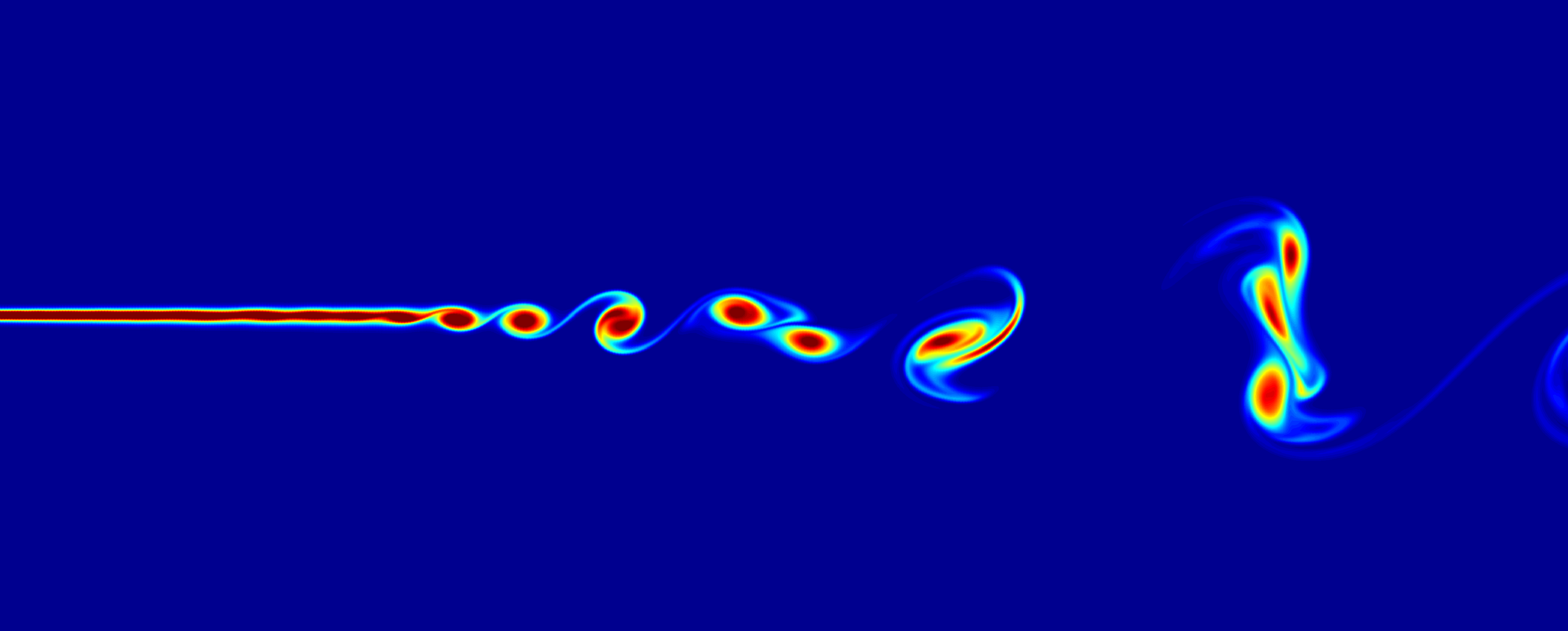};
        \end{axis}
    \end{tikzpicture}
\caption{Vorticity contours from a DNS of the mixing at $\mathop{\rm Re_{\delta_\omega}}=500$\label{fig:shear_snapshot}}
\end{figure}
\subsubsection{POD basis functions of the mixing layer}
\label{sec:mixing layer modes}
\begin{figure}
    \centering
            \subfigure[$\bm{u}_1$\label{sufig:shear_w_1}]
                {
                \begin{tikzpicture}
                    \begin{axis}[
                    ybar, xbar,
                        axis on top,
                        enlargelimits=false,
                        width=8cm,
                        height=3.5cm,
                        xlabel={$x$},
                        ylabel={$y$},
                        xtick={0,50,100,150,200,250,300},
                        ytick={-50,0,50},
                        ]
                            \addplot graphics
                            [xmin=0,xmax=300,ymin=-50,ymax=50,
                            includegraphics={trim=0 0 0 0,clip},]
                            {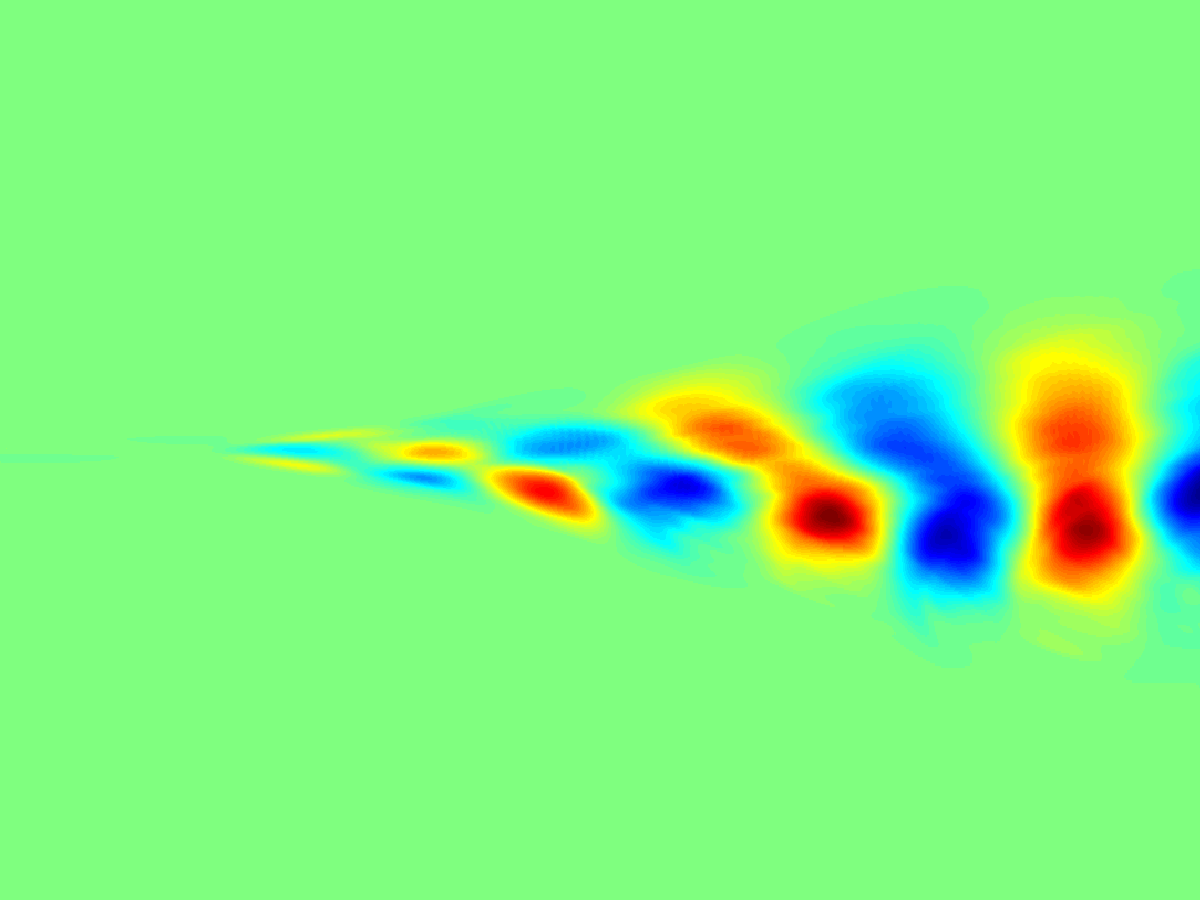};
                    \end{axis}
                \end{tikzpicture}
                }
            \subfigure[$\bm{u}_2$\label{sufig:shear_w_2}]
                {
                \begin{tikzpicture}
                    \begin{axis}[
                    ybar, xbar,
                        axis on top,
                        enlargelimits=false,
                        width=8cm,
                        height=3.5cm,
                        xlabel={$x$},
                        ylabel={$y$},
                        xtick={0,50,100,150,200,250,300},
                        ytick={-50,0,50},
                        ]
                            \addplot graphics
                            [xmin=0,xmax=300,ymin=-50,ymax=50,
                            includegraphics={trim=0 0 0 0,clip},]
                            {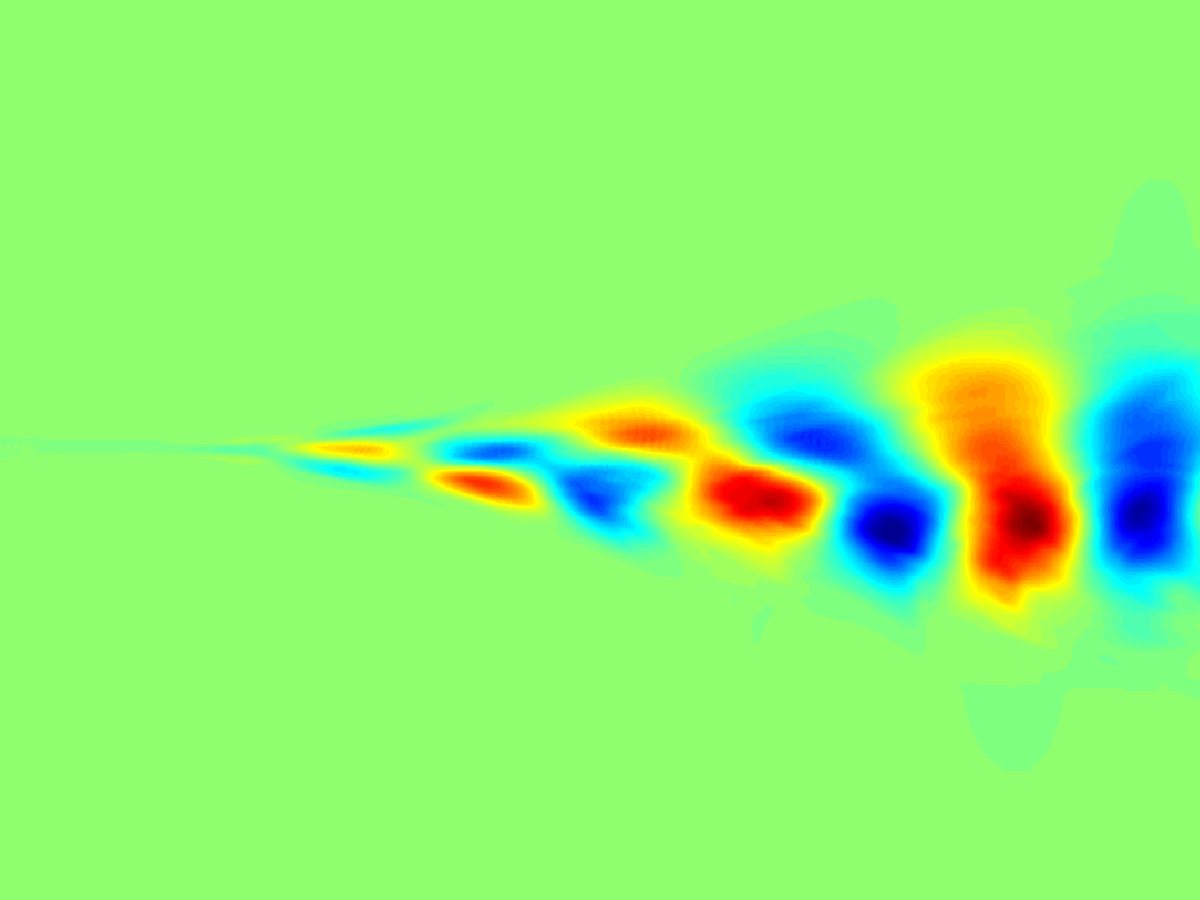};
                    \end{axis}
                \end{tikzpicture}
                }
            \subfigure[$\bm{u}_{20}$\label{sufig:shear_w_20}]
                {
                \begin{tikzpicture}
                    \begin{axis}[
                    ybar, xbar,
                        axis on top,
                        enlargelimits=false,
                        width=8cm,
                        height=3.5cm,
                        xlabel={$x$},
                        ylabel={$y$},
                        xtick={0,50,100,150,200,250,300},
                        ytick={-50,0,50},
                        ]
                            \addplot graphics
                            [xmin=0,xmax=300,ymin=-50,ymax=50,
                            includegraphics={trim=0 0 0 0,clip},]
                            {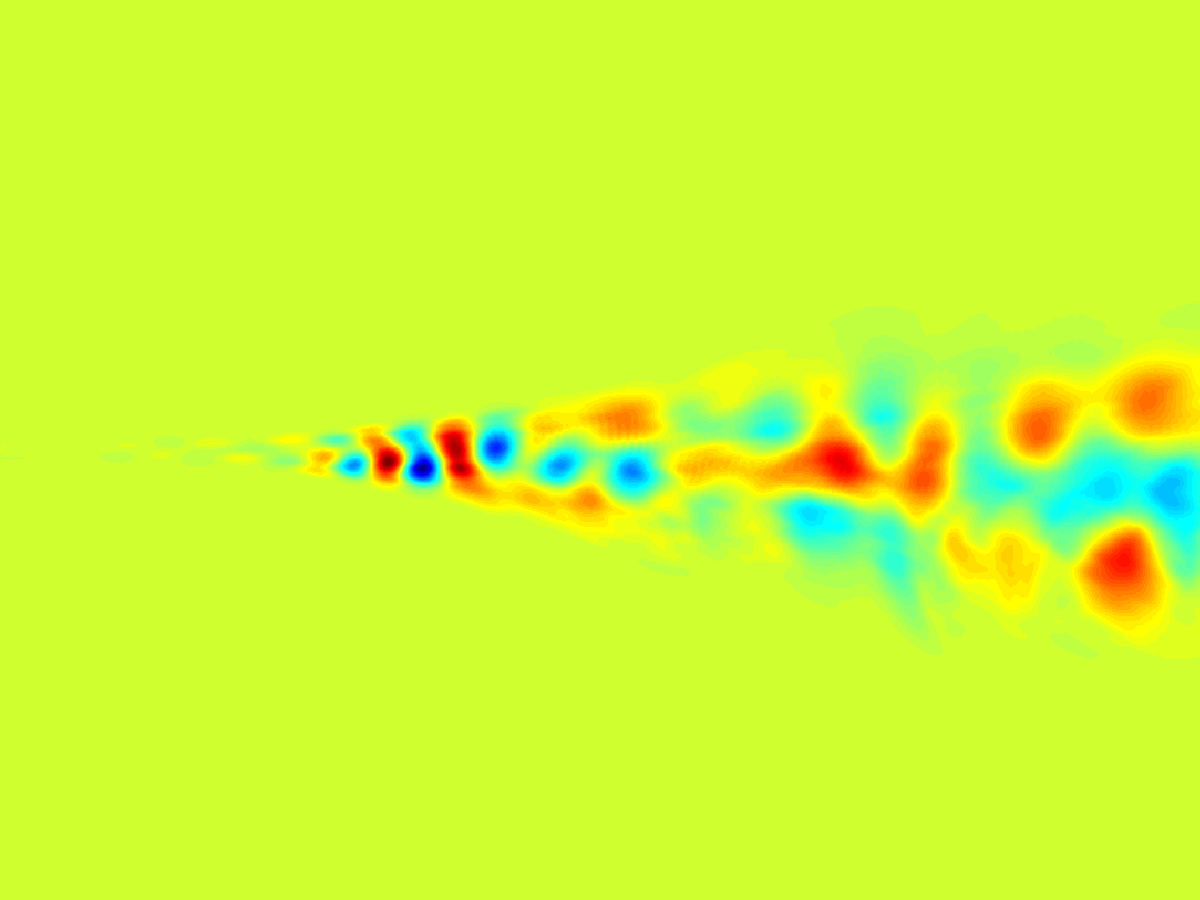};
                    \end{axis}
                \end{tikzpicture}
                }
            \subfigure[$\bm{u}_{100}$\label{sufig:shear_w_150}]
                {
                \begin{tikzpicture}
                    \begin{axis}[
                    ybar, xbar,
                        axis on top,
                        enlargelimits=false,
                        width=8cm,
                        height=3.5cm,
                        xlabel={$x$},
                        ylabel={$y$},
                        xtick={0,50,100,150,200,250,300},
                        ytick={-50,0,50},
                        ]
                            \addplot graphics
                            [xmin=0,xmax=300,ymin=-50,ymax=50,
                            includegraphics={trim=0 0 0 0,clip},]
                            {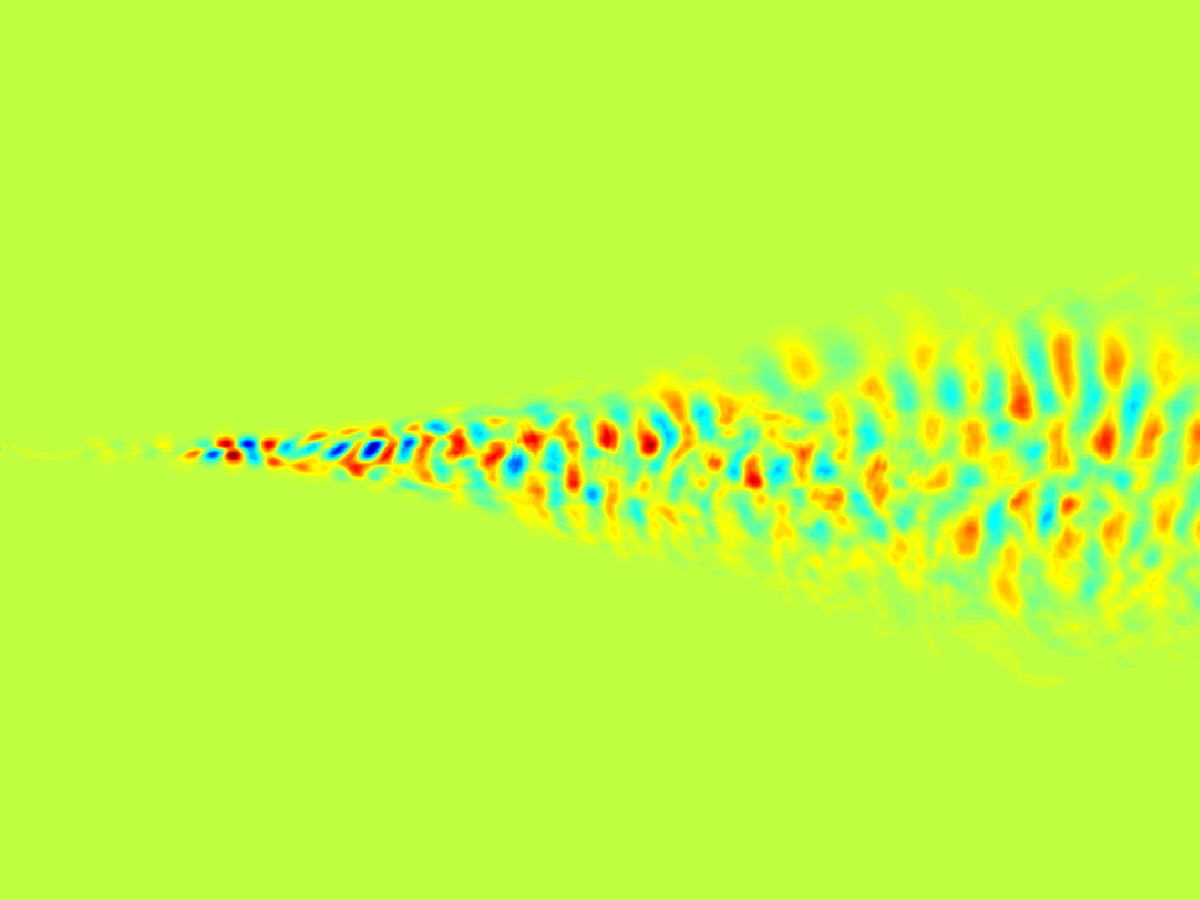};
                    \end{axis}
                \end{tikzpicture}
            }
    \caption{Vorticity contours of spatial POD basis functions of the mixing layer at $\mathop{\rm Re_{\delta_\omega}}=500$.}\label{fig:shear_POD_spatial_basis_functions}
    \end{figure}

A database of 2000 DNS snapshots of the mixing-layer cavity were used to find the POD basis functions, $\bm{u}_i(\bm{x})$ and $a_i(t)$. The nondimensional time interval between each snapshot equaled 1. Increasing the number of snapshots or the time interval between each snapshot had no significant effect on the performance characteristics of the ROMs. The percentaged of time-averaged, turbulent kinetic energy captured by the first $n$ POD basis functions of the mixing layer are summarized in Table~\ref{tab:lid_eigenvalues}.
\begin{table}
    \begin{center}
    \def~{\hphantom{0}}
        \begin{tabular}{c c}
            n & $E^{[1, \ldots ,n]}$ \\
            1 & 18.05 \\
            2 & 34.69 \\
            3 & 40.20 \\
            4 & 4.62 \\
            5 & 50.71 \\
            10 & 66.80 \\
            20 & 81.77 \\
            50 & 93.93 \\
            100 & 98.36 \\
        \end{tabular}
    \end{center}
    \caption{Percent of time-averaged, turbulent kinetic energy captured by the first $n$ POD basis functions of the mixing layer at $\mathop{\rm Re_{\delta_\omega}}=500$.}\label{tab:shear_eigenvalues}
\end{table}
Vorticity contours of spatial POD basis functions, $\bm{u}_i(\bm{x})$ for $i=1,2,20$ and $100$ of the mixing layer are illustrated in Fig.~\ref{fig:shear_POD_spatial_basis_functions}. As expected, the overall spatial wavelengths of the POD basis functions tends to increases with order. The low-order POD basis functions correspond to the large, high-energy physical scales of the turbulent flows while the higher-order POD basis functions correspond to the small, low-energy physical scales of the turbulent flow.
\subsubsection{Galerkin ROMs of mixing layer using standard POD basis functions, $a_i(t)$ and $\bm{u}_i(\bm{x})$}
\label{sec:Mixing_layer_standard_ROMS}
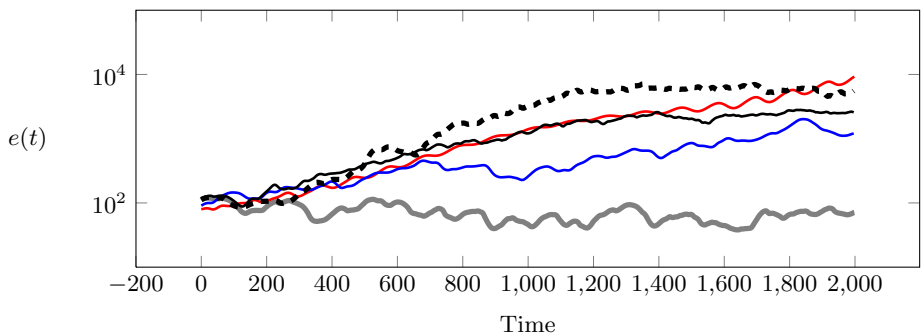
\begin{figure}
\centering
    \begin{tikzpicture}
        \begin{semilogyaxis}[
            ymin=10,
            ymax=1e5,
            width=12cm,
            height=5cm,
            xlabel={Time},
            ylabel={ $e(t)$ },
            y label style={at={(-0.02,0.5)}},
            legend columns=1,
            ]
            \addplot+[line join=round][
                line width=2pt,
                color=gray,
                mark={false},
                ]
                file {data/SHEAR_E_DNS.dat}; \label{legend:thick_gray}
            \addplot+[line join=round][
                line width=1pt,
                color=red,
                mark={false},
                ]
                file {data/SHEAR_E_ROM_0005_0005_0.dat};
            \addplot+[line join=round][
                line width=1pt,
                color=blue,
                mark={false},
                ]
                file {data/SHEAR_E_ROM_Standard_0010.dat};
            \addplot+[line join=round][
                line width=2pt,
                color=black,
                dashed,
                mark={false},
                ]
                file {data/SHEAR_E_ROM_Standard_0050.dat};
            \addplot+[line join=round][
                line width=1pt,
                color=black,
                mark={false},
                ]
                file {data/SHEAR_E_ROM_Standard_0100.dat};
         \end{semilogyaxis}
    \end{tikzpicture}
\caption{Galerkin-based ROMs of the mixing layer derived using the first $n=5$(\ref{legend:thin_red}), $10$(\ref{legend:thin_blue}), $50$(\ref{legend:dashed_black}) and $100$(\ref{legend:thin_black}) POD basis functions; DNS (\ref{legend:thick_gray})}\label{fig:SHEAR_e_STANDARD}
\end{figure}

\begin{figure}
    \centering
            \begin{tikzpicture}
                \begin{axis}[
                    ymin=-0.02,
                    ymax=0.08,
                    width=9cm,
                    height=4cm,
                    xlabel={n},
                    ylabel={ $\left\langle {\frac{d}{{dt}}e^{[1, \ldots ,n]}(t) } \right\rangle_T$ },
                    y label style={at={(-0.05,0.5)}},
                    legend columns=1,
                    ]
                    \addplot+[line join=round][
                        line width=1pt,
                        color=black,
                        mark={false},
                        ]
                        file {data/SHEAR_rate_of_e.dat};
                 \end{axis}
            \end{tikzpicture}
    \caption{Time-averaged rate of turbulent kinetic energy production associated with the first $n$ POD basis functions of the mixing layer}\label{fig:SHEAR_rate_of_e}
\end{figure}
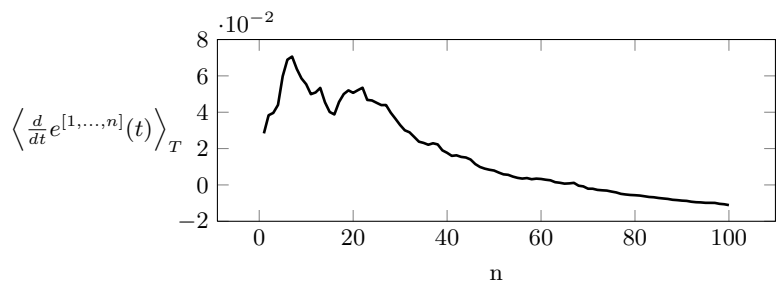

In this section, Galerkin ROMs of the mixing layer are derived using the standard POD basis functions, $a_i(t)$ and $\bm{u}_i(\bm{x})$ for $i=1,\cdots,n$. The turbulent kinetic energy as predicted by these ROMs is illustrated in Figure~\ref{fig:SHEAR_e_STANDARD}. Similarly to the lid-driven cavity test case, Galerkin ROMs of the mixing layer based on POD basis function tend to over predict the kinetic energy of the flow. 
Similarly to the lid-driven cavity case, these inaccuracies were anticipated. As illustrated in Figure~\ref{fig:SHEAR_rate_of_e}, the rate of kinetic energy production associated with the POD basis functions is positive.
\subsubsection{Galerkin ROMs of the mixing layer using the new basis functions, ${\tilde a_i(t)}$ and ${\tilde {\bm{u}}_i(\bm{x})}$}
\label{sec:Mixing_layer_new_ROM}
\begin{figure}
\centering
    \begin{tikzpicture}
        \begin{semilogyaxis}[
            ymin=10,
            ymax=1e5,
            width=12cm,
            height=5cm,
            xlabel={Time},
            ylabel={ $e(t)$ },
            y label style={at={(-0.02,0.5)}},
            legend columns=1,
            ]
            \addplot+[line join=round][
                line width=2pt,
                color=gray,
                mark={false},
                ]
                file {data/SHEAR_E_DNS.dat};
            \addplot+[line join=round][
                line width=2pt,
                color=black,
                dashed,
                mark={false},
                ]
                file {data/SHEAR_E_ROM_0005_0005_0.dat};
            \addplot+[line join=round][
                line width=1pt,
                color=black,
                mark={false},
                ]
                file {data/SHEAR_E_ROM_0005_0010_1.dat};
                \draw (axis cs:2000,1e4) node[left]
                {$n = 5$};
         \end{semilogyaxis}
    \end{tikzpicture}
    \begin{tikzpicture}
        \begin{semilogyaxis}[
            ymin=10,
            ymax=1e5,
            width=12cm,
            height=5cm,
            xlabel={Time},
            ylabel={ $e(t)$ },
            y label style={at={(-0.02,0.5)}},
            legend columns=1,
            ]
            \addplot+[line join=round][
                line width=2pt,
                color=gray,
                mark={false},
                ]
                file {data/SHEAR_E_DNS.dat};
            \addplot+[line join=round][
                line width=2pt,
                color=black,
                dashed,
                mark={false},
                ]
                file {data/SHEAR_E_ROM_0010_0010_0.dat};
            \addplot+[line join=round][
                line width=1pt,
                color=black,
                mark={false},
                ]
                file {data/SHEAR_E_ROM_0010_0020_1.dat};
                \draw (axis cs:2000,1e4) node[left]
                {$n = 10$};
         \end{semilogyaxis}
    \end{tikzpicture}
    \begin{tikzpicture}
        \begin{semilogyaxis}[
            ymin=10,
            ymax=1e5,
            width=12cm,
            height=5cm,
            xlabel={Time},
            ylabel={ $e(t)$ },
            y label style={at={(-0.02,0.5)}},
            legend columns=1,
            ]
            \addplot+[line join=round][
                line width=2pt,
                color=gray,
                mark={false},
                ]
                file {data/SHEAR_E_DNS.dat};  \label{legend:thick_gray};
            \addplot+[line join=round][
                line width=2pt,
                color=black,
                dashed,
                mark={false},
                ]
                file {data/SHEAR_E_ROM_0020_0020_0.dat};
            \addplot+[line join=round][
                line width=1pt,
                color=black,
                mark={false},
                ]
                file {data/SHEAR_E_ROM_0021_0040_1.dat};
                \draw (axis cs:2000,1e4) node[left]
                {$n = 20$};
         \end{semilogyaxis}
    \end{tikzpicture}
\caption{Evolution of the turbulent kinetic energy of the mixing layer as predicted by the DNS (\ref{legend:thick_gray}), a standard POD/Galerkin-based ROM (\ref{legend:dashed_black}) and the new Galerkin ROM (\ref{legend:thin_black}).}\label{fig:SHEAR_e}
\end{figure}
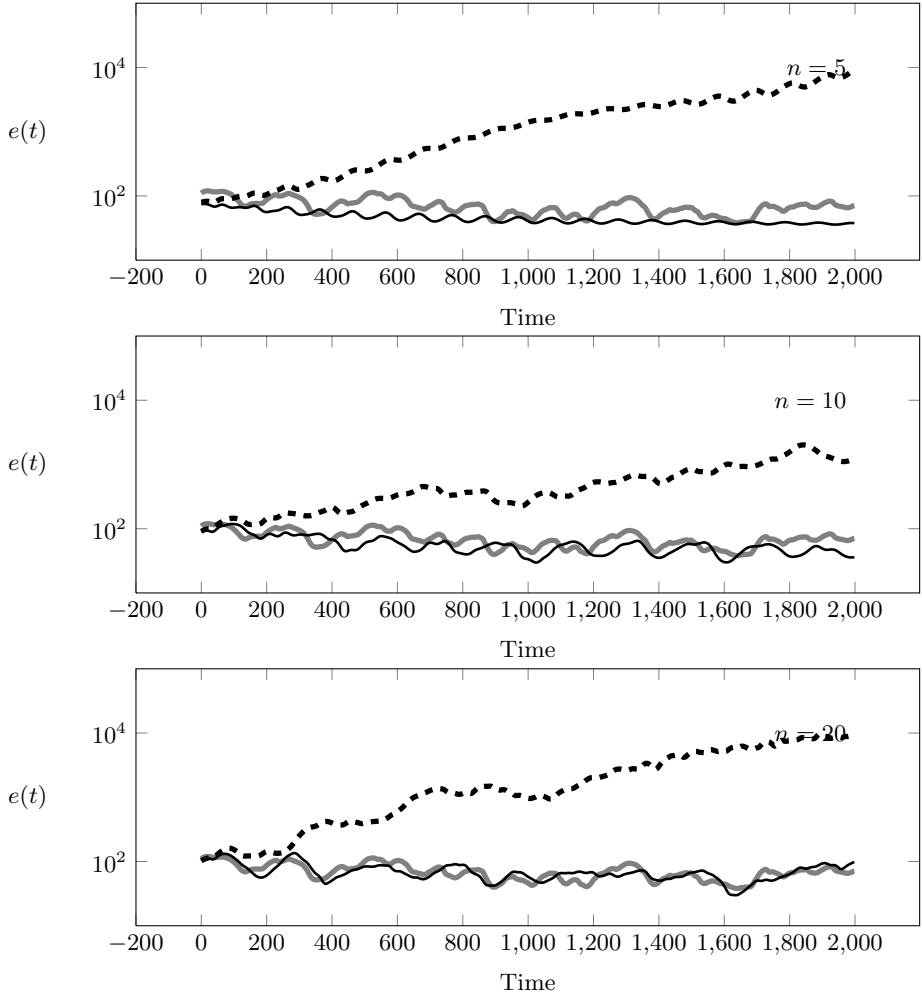

\indent In this section, the proposed new methodology is used to derive more accurate Galerkin ROMs of the mixing layer. Figure~\ref{fig:SHEAR_e} illustrates the evolution of the turbulent kinetic energy of the mixing layer as predicted by the new Galerkin ROMs. For the sake of brevity, only $n=5$, 10 and 20 ROMs are illustrated. The turbulent kinetic energy of the DNS solution is identified by the thick grey curves while the dashed and solid black curves identify the turbulent kinetic energy of the standard and new ROMs respectively. As expected, the Galerkin-based ROMs of the lid-driven cavity derived using standard POD basis functions over predict the kinetic energy of the flow. On the other hand, ROMs derived using the new basis functions converge to the correct mean of kinetic energy of DNS. As stated previously, these new basis functions have negative rates of kinetic energy production, $\epsilon$ while the standard POD basis functions have positive rates of kinetic energy production. Despite this difference, these new basis functions remain very similar to the POD basis functions as summarized in table~\ref{tab:SHEAR_e of modes}.

\begin{table}
    \begin{center}
    \def~{\hphantom{0}}
        \begin{tabular}{l c c}
            n & $E^{[1, \ldots ,n]}$ &  $\tilde E^{[1, \ldots ,n]}$ \\
            5 & 50.71 & 50.21 \\
            10 & 66.80 & 66.67\\
            20 & 81.77 & 79.14\\
        \end{tabular}
    \end{center}
    \caption{Percentage of time-averaged, turbulent kinetic energy captured by the first $n$ basis functions of the mixing layer.}\label{tab:SHEAR_e of modes}
\end{table}
\section{Conclusions}
\label{sec:Conclusions}
A new approach to Model Order Reduction of the Navier-Stokes Equations was proposed. The method provides spatial basis functions different from the usual proper orthogonal decomposition (POD) basis functions. These new basis functions resolve a greater range of physical scales of the turbulent fluid flow compared to the POD basis functions that are, by construction, biased toward the large, energy containing scales. When these basis functions are used in a Galerkin projection of the Navier-Stokes equations, more stable and accurate ROMs are derived.
\section*{Acknowledgements}
\label{sec:Acknowledgements}
The authors acknowledge financial support from Natural Science and Engineering Research Council of Canada (NSERC) and the ANR Chair of Excellence TUCOROM. The authors would like acknowledge Charbel Farhat and David Amsallem of Stanford Univeristy and Laurent Cordier, Joel Delville and Vladimir Parezanovicand from the institute Pprime in Poitiers, France for their invaluable comments and suggestions regarding the propose methodology. The authors are particularly grateful to Guillame Daviller for making available to use the mixing layer data set.

\clearpage
\appendix
\section{}\label{appA}
The following are MATLAB functions that solve the minimization problem for a given \verb=delta_e=. The inputs of the function \verb=stabilizeROM= are the \verb=N= spatial basis functions, \verb=u= and the temporal basis funcitons, \verb=a=. The function \verb=stabilizeROM= outputs the transformation matrix, \verb=X= and matrices containing the new spatial basis functions, \verb=u_tilde= and temporal basis functions \verb=a_tilde=. The objective function, \verb=objective= evaluates the difference between the optimal reconstruction of the time-averaged, turbulent kinetic energy using the POD basis functions and the new basis functions. The constraint function \verb=constraint= functions evaluates the time-averaged, turbulent kinetic energy.
\begin{lstlisting}
function [X,u_tilde,a_tilde] = stabilizeROM(u,a,L,N,n)
    global lambda L delta_e

    for i=1:N;
        lambda(i) = mean(a(i,:).*a(i,:));
    end

    x0 = eye(n,n);
    x0(N,:) = 0;

    problem = createOptimProblem('fmincon',...
                'objective', @objective, ...
                'nonlcon', @constraint, ...
                'x0',x0);
    [x,fval,EXITFLAG,OUTPUT,LAMBDA] = fmincon(problem)
    OUTPUT.message

    X=x*(x'*x)^(-1/2);
    u_tilde=u*X;
    a_tilde=X'*a;
end
\end{lstlisting}

\begin{lstlisting}
function [f] = objective(x)
global lambda;
    X = x*(x'*x)^(-1/2);
    sum_lambda = sum(lambda);
    lambda_tilda = diag(X'*diag(lambda)*X);
    sum_lambda_tilde = sum(lambda_tilde);

    f = sum_lambda - sum_lambda_tilde;
end
\end{lstlisting}

\begin{lstlisting}
function [c,ceq] = constraint(x,delta_e)
global D lambda;
    X=x*(x'*x)^(-1/2);
    L_tilde = X'*L*X;
    lambda_tilde = (X'*diag(lambda)*X);
    ceq = sum(sum(D_tilde.*lambda_tilde)) - delta_e;
    c = [];
end
\end{lstlisting}

\bibliographystyle{jfm}

\bibliography{data/References}

\end{document}